\documentclass[showpacs,preprintnumbers,amsmath,amssymb]{revtex4}

\usepackage{graphicx}
\usepackage{dcolumn}
\usepackage{bm}

\begin{document}


\title{Scattering states of coupled valence-band holes in point defect potential derived from variable phase theory}

\author{P. Bogdanski}
\affiliation{D\'epartement de Physique-EEA, UFR de Sciences, Universit\'e de Caen, Boulevard Mar\'echal Juin, 14032 Caen Cedex, France}
\email{Patrick.Bogdanski@physique-eea.unicaen.fr}
\author{H. Ouerdane}%
\affiliation{Laboratoire CRISMAT UMR CNRS-ENSICAEN 6508, 6 Boulevard Mar\'echal Juin, 14050 Caen Cedex, France}

\date{\today}

\begin{abstract}
In this article we present a method to compute the scattering states of holes in spherical bands in the strong spin-orbit coupling regime. More precisely, we calculate scattering phase shifts and amplitudes of holes induced by defects in a semiconductor crystal. We follow a previous work done on this topic by Ralph [H. I. Ralph, Philips Res. Rept. {\bf 32} 160 (1977)] to account for the $p$-wave nature and the coupling of valence band states. We extend Ralph's analysis to incorporate finite-range potentials in the scattering problem. We find that the variable phase method provides a very convenient framework for our purposes and show in detail how scattering amplitudes and phase shifts are obtained. The Green's matrix of the Schr\"odinger equation, the Lippmann-Schwinger equation and the Born approximation are also discussed. Examples are provided to illustrate our calculations with Yukawa type potentials.
\end{abstract}

\maketitle
\section{Introduction}

The valence bands of the semiconductors of the column IV (Si, Ge) and of some III-V (GaAs, InP) as well as II-VI (CdTe, ZnSe) compounds have a similar structure. Because of the complexity of this structure, calculation of the drift mobility $\mu_{\rm d}$ and the mobility of Hall $\mu_{\rm H}$ in these materials, when they are of the p-type, is a very difficult problem. Simplifying hypothesis usually consist of neglecting the split-off band and assuming parabolic band dispersions, $E_i({\bf k})$, for the light ($i = \ell$), and heavy ($i = {\rm h}$) hole bands. Even in these conditions, calculations remain complicated since one has to take into account the $p$-wave nature of the holes wavefunctions, as well as the various possible transitions between the two subbands.

In the semi-classical model, the evaluation of the transport coefficients requires the knowledge of the non-equilibrium hole occupation functions $f_i({\bf k})$ ($i = \ell, {\rm h}$), solutions of the Boltzmann equation. The transition rates between the states $|i, {\bf k}\rangle$ and $|j, {\bf k}'\rangle$, characterizing the various hole scattering processes (phonons, defects $\cdots$) are included in the definition of the collision integral \cite{COS74,SZM86}. In the present work, we focus on the problem of elastic scattering by a point defect, modeled by a spherically symetric perturbation potential.

In a weak field regime, the Boltzmann equation can be linearized and the scattering by the defects are characterized by four relaxation times, $\tau_{ij}$, that correspond to the four possible transitions within (intra) and between (inter) the two bands $i$ and $j$. In the Born approximation, the various $\tau_{ij}$ are calculated introducing the overlap factors $G_{ij}({\bf k},{\bf k}')$ \cite{EHR59,WIL71,NAG80} that characterize the $p$-wave nature of the hole wavefunctions. For the Yukawa potential, frequently used to model ionized defects, expressions of the times $\tau_{ij}$ are well known \cite{BIR62,COS73}. In the low temperature regime, where the scattering by defects become dominant compared to the other scattering mechanisms, the thermal energy of the holes is small compared to the perturbation induced by the defects in the crystal, and the Born approximation ceases to be valid.

To overcome this problem, the phase shift method which yields an essentially exact solution to scattering problems, can be used. In the case of carriers that belong to the same band (no coupling) of $s$-wave nature, the expression of the relaxation time $\tau$ as function of the phase shifts $\delta_l$ of each partial wave with angular momentum $l$, is well known \cite{HUA48}. Meyer and Bartoli have used this result to perform approximate calculations of carrier mobility limited by scattering on ionized impurities in p-doped Si and GaAs \cite{Meyer}. In their model they neglected the interband transitions and ignored the $p$-wave nature of the holes wavefunctions. They considered a Yukawa potential and gave approximate analytic forms allowing direct calculation of mobilities without the need to compute the phase shifts.

Kim and Majerfeld have calculated phase shifts to compute the scattering states of holes on a single ionized impurity \cite{Kim}. Their model has nonetheless two shortcomings: they assume a weak perturbation and their phase shifts calculations are performed using a Hamiltonian different from the well established one usually found in many articles reporting on the defects in p-type semiconductors \cite{Ralph,Baldereschi1,JOO78,SZM86,SAI87}.

A more rigorous treatment of the scattering of holes based on the phase shift method was done by Ralph \cite{Ralph}. Calculations were done with a spherical Hamiltonian in the strong spin-orbit coupling regime. For each of the two regular independent solutions of the Schr\"odinger equation, the radial functions associated to each scattering partial wave were completely characterized by two amplitudes, $a^{\pm}$, and two phase shifts, $\delta^{\pm}$, in the asymptotic region ($r\rightarrow \infty$); $a^{\pm}$ and $\delta^{\pm}$ are called Ralph's parameters in the present article. In this model the contribution of each partial wave to the various scattering times $\tau_{ij}$ thus required the calculation of eight parameters: four scattering amplitudes and four scattering phase shifts. The example treated by Ralph was only an academic one since a zero-range potential of the Dirac-type was used, for which it is possible to obtain simple analytic expressions of the $\tau_{ij}$. Ralph's parameters had never been computed for more realistic potentials, such as a Yukawa potential, much used for the study of electron and hole scattering by ionized defects.

In this article we show how to make use of the variable phase method \cite{Calogero} to obtain Ralph's parameters for finite range potentials. The variable phase method has been mostly developped and used for collisions and scattering problems in atomic and nuclear physics, see e.g. \cite{MorseAllis,DRU49,BAB65,LED77,BER86,BAL98,henni1, henni2}, whereas there are comparatively few works where it is used in solid state physics, e.g. \cite{portnoi2,portnoi3}. To the best of our knowledge this method had never been applied to scattering problems with coupling in solid state physics. In its simplest version (no coupling) the variable phase approach consists of deriving a first-order differential equation satisfied by the phase shifts $\delta_l(r)$ of the partial waves with angular momentum $l$ induced by a potential truncated at distance $r$, and solve it imposing $\delta_l(0) = 0$ as an initial condition. The method is also useful to calculate the amplitudes $a_l(r)$ of the partial waves. For the mixed valence-band holes the present work consists of tranforming a 2$\times$2 second order differential system satisfied by the two radial wavefunctions into a 4$\times$4 first order differential system whose solutions are the two amplitudes $a^{\pm}_l(r)$ and the two phase shifts $\delta^{\pm}_l(r)$.

The article is organised as follows: in Section II, the spherical model Hamiltonian on which we base all the subsequent calculations is introduced. In Section III we establish the framework of our calculations. We show how we derive the generalized coupled first-order differential equations satisfied by the scattering phase shifts and amplitudes of heavy and light hole wavefunctions. The Green's matrix of the Schr\"odinger equation, the Lippmann-Schwinger equation and the Born approximation are discussed in detail in section IV. The object of Section V is the study of the non-trivial behavior of the wavefunctions near the origin, the knowledge of which is crucial for successful numerical computation of the solutions of the abovementioned differential equations. In section VI we give examples based on Yukawa-type potentials to illustrate our calculations before concluding.

\section{Spherical model Hamiltonian in the strong coupling regime}

Luttinger showed that in the strong coupling regime the holes can be seen as quasi-particles of spin $J=3/2$ \cite{LUT56}. Baldereschi and Lipari treated the problem of acceptor states in the effective mass approximation in order to separate the symmetrically spherical and cubic terms in the Hamiltonian \cite{Baldereschi1}. They showed that the spherical Hamiltonian, ${\mathcal H}_{\rm s}$, may be written as:

\begin{equation}\label{eq1}
{\mathcal H}_{\rm s}({\bf r},{\bf p}) = \frac{\displaystyle \gamma_1}{\displaystyle 2m_0}~{\bf p}^2 - \frac{\displaystyle \gamma_1}{\displaystyle 18m_0}~\mu~  P^{(2)}\cdot J^{(2)} + V(r),
\end{equation}

\noindent where $m_0$ is the free electron mass, $\mu = (4\gamma_2 +6\gamma_3)/5\gamma_1$, and $\gamma_1$, $\gamma_2$ and $\gamma_3$ are the Luttinger parameters \cite{LUT56}. The first term of ${\mathcal H}_{\rm s}$ in Eq.~(\ref{eq1}) is the kinetic energy, the second term represents the spherical part of the spin-orbit interaction, and the last one is the perturbation potential created by the defect that is assumed to depend only on the distance $r$. The diagonalization of the unperturbed Hamiltonian, ${\mathcal H}_{\rm s}^0({\bf p})$, where ${\bf p} = \hbar{\bf k}$, yields two parabolic bands characterized by the effective mass coefficients $1/(1+\mu)\gamma_1 = r_{\rm p}^2/\gamma_1$ for the light holes and $1/(1-\mu)\gamma_1 = r_{\rm m}^2/\gamma_1$ for the heavy holes. Denoting $E({\bf k}) = \hbar^2 \gamma_1 {\bf k}^2/2m_0$ the hole energy which is a constant of motion, the wavevectors of the light and heavy holes are respectively $k^+ = r_{\rm p} k$ and $k^- = r_{\rm m} k$.

The Hamiltonian ${\mathcal H}_{\rm s}$ commutes with the total angular momentum ${\bf F} = {\bf L} +{\bf J}$, where ${\bf L}$ is the angular momentum. Therefore the scattering eigenstates may be written as $|3/2,F,M\rangle$, where $F(F+1)$ and $M$ are the eigenvalues of ${\bf F}^2$ and $F_z$ respectively.

For a given value of $F$ ($F \geq 3/2$) there are 4 corresponding values of $L$: $L=F-3/2$, $L=F-1/2$, $L=F+1/2$ and $L=F+3/2$. Moreover  since the parity $\Pi = (-1)^L$ is conserved, only the states $|L,3/2,F,M\rangle$ satisfying $\Delta L = 0,\pm 2$ are coupled. In the sub-space $|F,M\rangle$, the Hamiltonian matrix ${\mathcal H}_{\rm s}$ contains two $2\times 2$ blocks of well defined parity. The first block is associated to the states $\{|L=F-3/2\rangle,|L=F+1/2\rangle\}$, and the second to the states $\{|L=F-1/2\rangle,|L=F+3/2\rangle\}$. From now on we denote $|L-1\rangle$ and $|L+1\rangle$ the two coupled states of parity $\Pi$ and we shall use the parameter $\xi = \pm 1$ introduced by Ralph \cite{Ralph} to distinguish the value of $L$ corresponding to each block: if $\xi = 1$, $L=F-1/2$ and if $\xi=-1$, $L=F+1/2$.

The two radial wavefunctions $f_{L\pm 1}(r)$ associated with the scattering states $|F,\xi,M\rangle$ satisfy a system of coupled radial Schr\"odinger equations, which reads \cite{Baldereschi1}:

\begin{eqnarray}\label{eq4}
\left(\begin{array}{cc}
H_{L-1,L-1} & H_{L-1,L+1}\\
H_{L+1,L-1} & H_{L+1,L+1}\\
\end{array}\right)
\left(\begin{array}{c}
f_{L-1}\\
f_{L+1}\\
\end{array}\right)
=E
\left(\begin{array}{c}
f_{L-1}\\
f_{L+1}\\
\end{array}\right),
\end{eqnarray}

\noindent where the four matrix elements $H_{L\pm 1,L\pm 1}$ are second order differential operators of the variable $r$. The perturbation potential $V(r)$ appears only on the diagonal of the Hamiltonian matrix.

\begin{table}
\caption{\label{tab:table1} Values of $\cos\alpha$ and $\sin\alpha$ as function of $\xi$ and $L$.}
\begin{ruledtabular}
\begin{tabular}{cc}
$\rule{0pt}{5ex}\xi = -1, L=F+\frac{\displaystyle 1}{\displaystyle 2}, L \ge 2$ & $\xi = 1, L=F-\frac{\displaystyle 1}{\displaystyle 2}, L \ge 1$ \vspace{0.25cm}\\ \hline
$\rule{0pt}{5ex}\cos\alpha = -\frac{\displaystyle L+2}{\displaystyle 2L+1}$ & $\cos\alpha = \frac{\displaystyle L-1}{\displaystyle 2L+1}$  \\
$\rule{0pt}{5ex}\sin\alpha = \frac{\displaystyle \sqrt{3(L+1)(L-1)}}{\displaystyle 2L+1}$ & $\sin\alpha = \frac{\displaystyle \sqrt{3L(L+2)}}{\displaystyle 2L+1}$\vspace{0.2cm}\\
\end{tabular}
\end{ruledtabular}
\end{table}

\section{Derivation of the phase equation}

Ralph has demonstrated that in a region where the perturbating potential has no effect, the two coupled radial wavefunctions $f_{L\pm 1}(r)$ may be characterized by four constant parameters: two amplitudes $a^{\pm}$ and two phase shifts $\delta^{\pm}$ \cite{Ralph}. In this section we show how we obtain those four parameters as the limits, when $r \rightarrow \infty$, of the functions $a^{\pm}(r)$ and $\delta^{\pm}(r)$ associated to a spherically symmetric potential truncated at distance $r$.

\subsection{From 2$\times$2 second order differential system to 4$\times$4 first order differential system}

To eliminate the factor $r^{-1}$ in the solutions of the unperturbed system and in their asymptotic expressions, we define the functions $u_{L\pm 1}(r) = rf_{L\pm 1}(r)$. Introducing the dimensionless variable $x=kr$ and the reduced potential $v(x)=V/E$, and factorizing the differential operators $H_{L\pm 1,L\pm 1}$, we find that $u_{L\pm 1}(x)$ satisfy the following system:

\begin{equation}\label{eq9}
\left(\begin{array}{cc}
\left(1+\mu\cos\alpha\right)\left[\frac{\displaystyle {\rm d}}{\displaystyle {\rm d}x}+\frac{\displaystyle L}{\displaystyle x}\right] & -\mu\sin\alpha\left[\frac{\displaystyle {\rm d}}{\displaystyle {\rm d}x}+\frac{\displaystyle L}{\displaystyle x}\right]\\
\rule{0pt}{4ex}-\mu\sin\alpha\left[\frac{\displaystyle {\rm d}}{\displaystyle {\rm d}x}-\frac{\displaystyle L+1}{\displaystyle x}\right] & \left(1-\mu\cos\alpha\right)\left[\frac{\displaystyle {\rm d}}{\displaystyle {\rm d}x}-\frac{\displaystyle L+1}{\displaystyle x}\right]
\end{array}\right)
\left(\begin{array}{c}
\left[\frac{\displaystyle {\rm d}}{\displaystyle {\rm d}x}-\frac{\displaystyle L}{\displaystyle x}\right]u_{L-1}(x)\\
\rule{0pt}{4ex}\left[\frac{\displaystyle {\rm d}}{\displaystyle {\rm d}x}+\frac{\displaystyle L+1}{\displaystyle x}\right]u_{L+1}(x)\\
\end{array}\right)
=\left(v(x)-1\right)
\left(\begin{array}{c}
u_{L-1}(x)
\\
u_{L+1}(x)\\
\end{array}\right),
\end{equation}

\noindent where the angle $\alpha$ introduced by Gel'mont and D'yakonov \cite{Gelmont1} is defined for each value of $F$ and $\xi$ in Table 1.

To lower the order of this differential system and to obtain the free solutions ($v=0$) in a simpler way, we define the functions $G^{\pm}(x)$ as follows:

\begin{equation}\label{eq13}
\left(\begin{array}{c}
G^+\\
G^-\\
\end{array}\right)
= \left[R\right]^{-2}\left[\Omega\right]^{-1}
\left(\begin{array}{cc}
-1&0\\
\phantom{-}0&1
\end{array}\right)
\left(\begin{array}{c}
\left[\frac{\displaystyle {\rm d}}{\displaystyle {\rm d}x}-\frac{\displaystyle L}{\displaystyle x}\right]u_{L-1}(x)\\
\rule{0pt}{4ex}\left[\frac{\displaystyle {\rm d}}{\displaystyle {\rm d}x}+\frac{\displaystyle L+1}{\displaystyle x}\right]u_{L+1}(x)
\end{array}\right),
\end{equation}

\noindent where $\left[\Omega\right]$ is the 2$\times$ 2 rotation matrix of angle $\alpha/2$ and $\left[R\right]$ the diagonal matrix whose elements are $r_{\rm p}$ and $r_{\rm m}$.

Introducing the vectors $\vec{u}=\left(\begin{array}{c}u_{L-1}\\u_{L+1}\end{array}\right)$ and $\vec{G}=\left(\begin{array}{c}G^+\\G^-\end{array}\right)$, the system in Eq.(\ref{eq9}) take the form of a first order differential system perturbed by a $4\times 4$ matrix $\left[V(x)\right]$:

\begin{eqnarray}\label{eq16}
\frac{\displaystyle {\rm d}}{\displaystyle {\rm d}x}
\left(\begin{array}{c}
\vec{u}\\
\vec{G}\\
\end{array}\right)
= \left[A(x)\right]
\left(\begin{array}{c}
\vec{u}\\
\vec{G}\\
\end{array}\right)
-\left[V(x)\right]
\left(\begin{array}{c}
\vec{u}\\
\vec{G}\\
\end{array}\right),
\end{eqnarray}

\noindent which is a very convenient approach to study the solutions near the origin. In Eq.~(\ref{eq16}), $\left[A(x)\right]$ and $\left[V(x)\right]$ are defined as follows:

\begin{eqnarray}\label{eq17}
\left[A(x)\right] \!\!= \!\!\!
\left( \begin{array}{cc}
\frac{\displaystyle 1}{\displaystyle x}
\left( \begin{array}{c c}
L & 0\\
0 & -L-1\\
\end{array}\right) &
\left( \begin{array}{c c}
-1 & 0\\
0 & 1\\
\end{array}\right)\left[\Omega\right] \left[R\right]^2\\
\rule{0pt}{4ex}\left[\Omega\right]^{-1}
\left( \begin{array}{c c}
1 & 0\\
0 & -1\\
\end{array}\right)&\frac{\displaystyle \left[\Omega\right]^{-1}}{\displaystyle x}
\left( \begin{array}{c c}
-L & 0\\
0 & L+1\\
\end{array}\right)\left[\Omega\right]\!\!\!\\
\end{array}\right)
\end{eqnarray}

\noindent and

\begin{eqnarray}\label{eq18}
\left[V(x)\right]= v(x)
\left(\begin{array}{cc}
\left[O\right] & \left[O\right]\\
\rule{0pt}{4ex}\left[\Omega\right]^{-1}
\left( \begin{array}{c c}
1 & 0\\
0 & -1\\
\end{array}\right) & \left[O\right]\\
\end{array}\right),
\end{eqnarray}

\noindent where $\left[O\right]$ is a 2$\times$2 matrix whose elements are all zero.

\subsection{The phase equation}

When the potential $v(x)$ is zero, we find that the vector $\vec{G}$ satisfies a differential equation of the Schr\"odinger type:

\begin{equation}\label{eq19}
\frac{\displaystyle {\rm d}^2}{\displaystyle {\rm d}x^2}~ \vec{G} - \frac{\displaystyle L(L+1)}{\displaystyle x^2}~\vec{G}+\left[R\right]^2\vec{G} = 0.
\end{equation}

\noindent The above equation has four independent solutions of which two are regular solutions $\vec{G}^{\pm}_{0,{\rm r}}$ and two irregular $\vec{G}^{\pm}_{0,{\rm i}}$. The vectors $\vec{u}^+_{0}$ and $\vec{u}^-_{0}$, respectively associated to $\vec{G}_0^+$ and $\vec{G}_0^-$, may be readily obtained from Eq.~(\ref{eq16}) with $v=0$. It is convenient to put together the four free solutions, regular ({\rm r}) and irregular ({\rm i}) $\left(\begin{array}{c}\vec{u}\\ \vec{G} \\ \end{array}\right)^{\pm}_{0,{\rm r/i}}$ into a 4$\times$4 matrix $\left[W(x)\right]$ whose elements are given in Appendix A. The regular and irregular parts of this matrix may be expressed in quite a simple fashion with modified spherical Bessel ${\hat {\jmath}}_l$ and Neumann $\hat{n}_l$ functions respectively \cite{Calogero}. The general solution for $v(x)=0$ can thus be written:

\begin{eqnarray}\label{eq25}
\left(\begin{array}{c}
\vec{u}\\
\vec{G}\\
\end{array}\right)_0
=\left[W(x)\right]\vec{C}_0
\end{eqnarray}

\noindent where $\vec{C}_0$ is a four-component constant vector.

When the potential $v(x)$ is non-zero, we apply the Lagrange method of the variation of constants \cite{Ronveaux} to find the solutions:

\begin{eqnarray}\label{eq26}
\left( \begin{array}{c}
\vec{u}\\
\vec{G}\\
\end{array}\right)
=\left[W(x)\right] \vec{C}(x).
\end{eqnarray}

Since $\left[W(x)\right]$ is solution of the unperturbed system, i.e. ${\rm d}\left[W\right]/{\rm d}x = \left[A\right]\left[W\right]$, we find that the unknown vector $\vec{C}(x)$ satisfies the following differential equation:

\begin{eqnarray}\label{eq27}
\frac{\displaystyle {\rm d}\vec{C}}{\displaystyle {\rm d}x} = -\left[W\right]^{-1}\left[V\right]\left[W\right]\vec{C}=-\left[W\right]^{-1}\left[V\right]
\left( \begin{array}{c}
\vec{u}\\
\vec{G}\\
\end{array}\right).
\end{eqnarray}

In the framework of the variable phase method \cite{Calogero} the vector $\vec{C}(x)$ is searched in the form:

\begin{eqnarray}\label{eq32}
\vec{C} = 
\left(\begin{array}{c}
c^+(x)\\
-s^+(x)\\
c^-(x)\\
-s^-(x)\\
\end{array}\right),
\end{eqnarray}

\noindent where, for ease of notation, we have omitted the quantum numbers $(F,\xi,L)$.

Inserting Eq.~(\ref{eq32}) into Eq.~(\ref{eq27}) and performing the algebra yield the following system of differential equations:

\begin{subequations}
\begin{equation}\label{eq35a}
\frac{\displaystyle {\rm d}}{\displaystyle {\rm d}x}
\left(\begin{array} {c}
c^+(x)\\
s^+(x)\\
\end{array}\right)
=-v(x)
\left(\begin{array} {cc}
\cos\frac{\displaystyle \alpha}{\displaystyle 2}~\hat{n}_{L-1}(r_{\rm p}x) & \sin\frac{\displaystyle \alpha}{\displaystyle 2}~\hat{n}_{L+1}(r_{\rm p}x)\\
\rule{0pt}{4ex}\cos\frac{\displaystyle \alpha}{\displaystyle 2}~\hat{\jmath}_{L-1}(r_{\rm p}x) & \sin\frac{\displaystyle \alpha}{\displaystyle 2}~\hat{\jmath}_{L+1}(r_{\rm p}x)\\
\end{array}\right)
\left(\begin{array} {c}
u_{L-1}(x)\\
u_{L+1}(x)\\
\end{array}\right),
\end{equation}
\begin{equation}\label{eq35b}
\frac{\displaystyle {\rm d}}{\displaystyle {\rm d}x}
\left(\begin{array} {c}
c^-(x)\\
s^-(x)\\
\end{array}\right)
=-v(x)
\left(\begin{array} {cc}
-\sin\frac{\displaystyle \alpha}{\displaystyle 2}~\hat{n}_{L-1}(r_{\rm m}x) & \cos\frac{\displaystyle \alpha}{\displaystyle 2}~\hat{n}_{L+1}(r_{\rm m}x)\\
\rule{0pt}{4ex}-\sin\frac{\displaystyle \alpha}{\displaystyle 2}~\hat{\jmath}_{L-1}(r_{\rm m}x) & \cos\frac{\displaystyle \alpha}{\displaystyle 2}~\hat{\jmath}_{L+1}(r_{\rm m}x)\\
\end{array}\right)
\left(\begin{array} {c}
u_{L-1}(x)\\
u_{L+1}(x)\\
\end{array}\right).
\end{equation}
\end{subequations}

The variable phase shifts $\delta^{\pm}(x)$ and amplitudes $a^{\pm}(x)$ are now introduced in the same fashion as in problems without coupling \cite{Calogero}:

\begin{subequations}
\begin{equation}\label{eq36a}
c^{\pm}(x) = a^{\pm}(x)\cos\delta^{\pm}(x),
\end{equation}
\begin{equation}\label{eq36b}
s^{\pm}(x) = a^{\pm}(x)\sin\delta^{\pm}(x).
\end{equation}
\end{subequations}

Inserting the above expressions of $c^{\pm}$ and $s^{\pm}$ in the differential systems Eqs.~(\ref{eq35a}) and (\ref{eq35b}), we obtain the generalized non-linear coupled equations for the phases and amplitudes of the holes:

\begin{eqnarray}\label{eq37}
\frac{\displaystyle {\rm d}}{\displaystyle {\rm d}x}
\left(\begin{array} {c}
\delta^+\\
\delta^-\\
a^+\\
a^-\\
\end{array}\right)
=v(x)
\left(\begin{array} {c}
-r_{\rm p}D^+ + r_{\rm m}~a^-D_{\rm c}/a^+\\
-r_{\rm m}D^- + r_{\rm p}~a^+D_{\rm c}/a^-\\
-r_{\rm p}a^+SD^+ + r_{\rm m}a^-SD^{+,-}_{\rm c}\\
-r_{\rm m}a^-SD^- + r_{\rm p}a^+SD^{-,+}_{\rm c}\\
\end{array}\right),
\end{eqnarray}

\noindent where the functions $D^{\pm}$, $SD^{\pm}$, $D_{\rm c}$ and $SD^{\varepsilon,-\varepsilon}_{\rm c}$ ($\varepsilon \equiv \pm$) are given in Appendix B.

Since the wavefunctions $u_{L\pm 1}$ must be regular at the origin, the above differential system must be solved with $\delta^{\pm}(0) = 0$ as initial conditions. This choice is justified by the fact that if the potential is truncated at the origin, the perturbation is removed from the problem and hence the shifts of the phases of the wavefunctions are zero. However, the behavior of the functions $a^+$ and $a^-$ near the origin is not imposed. There are thus two degrees of freedom that allows one to generate two independent solutions. This point is discussed further in Section V.

\subsection{Asymptotic expressions of the wavefunctions}

We assume that the perturbation potential $V(r)$ has a finite range $r_0$ (the Coulomb potential is therefore excluded). For $x > kr_0$ and $x\gg L/r_{\nu}$ ($\nu \equiv {\rm p}, {\rm m}$) the functions $a^{\pm}(x)$ and $\delta^{\pm}(x)$ are practically constant:

\begin{subequations}
\begin{equation}\label{eq39a}
c^{\pm}(x) \approx a^{\pm}(\infty)\cos\delta^{\pm}(\infty) = a^{\pm}\cos\delta^{\pm}
\end{equation}
\begin{equation}\label{eq39b}
s^{\pm}(x) \approx a^{\pm}(\infty)\sin\delta^{\pm}(\infty) = a^{\pm}\sin\delta^{\pm},
\end{equation}
\end{subequations}

\noindent and the functions $\hat{\jmath}_l(r_{\nu}x)$ and $\hat{n}_l(r_{\nu}x)$ can be taken as circular functions: $\lim_{x\rightarrow\infty}\hat{\jmath}_l(r_{\nu}x) = \sin\left(r_{\nu}x-l\pi/2\right)$ and $\lim_{x\rightarrow\infty}\hat{n}_l(r_{\nu}x) = -\cos\left(r_{\nu}x-l\pi/2\right)$. Using Eq.~(\ref{eq26}) and formulas of Appendix A, we find the asymptotic expressions of the wave functions $u_{L\pm 1}$:

\begin{subequations}
\begin{equation}\label{eq40a}
u_{L-1}(x) \approx r_{\rm p}a^+\cos\frac{\displaystyle \alpha}{\displaystyle 2}~\cos\left(r_{\rm p}x - L\pi/2 + \delta^+\right) - r_{\rm m}a^-\sin\frac{\displaystyle \alpha}{\displaystyle 2}~\cos\left(r_{\rm m}x - L\pi/2 + \delta^-\right)
\end{equation}
\begin{equation}\label{eq40b}
u_{L+1}(x) \approx -r_{\rm p}a^+\sin\frac{\displaystyle \alpha}{\displaystyle 2}~\cos\left(r_{\rm p}x - L\pi/2 + \delta^+\right) - r_{\rm m}a^-\cos\frac{\displaystyle \alpha}{\displaystyle 2}~\cos\left(r_{\rm m}x - L\pi/2 + \delta^-\right).
\end{equation}
\end{subequations}

Comparing these expressions to those derived in Ref.~\cite{Ralph}, we see that the variable phase method is suitable to compute Ralph's parameters.

The functions $u_\ell$ and $u_h$ defined as linear combinations of $u_{L-1}$ and $u_{L+1}$:

\begin{eqnarray}\label{eq41}
\left(\begin{array}{c}
u_\ell(x)\\
u_h(x)
\end{array}\right)
= \left[\Omega\right]
\left(\begin{array}{c}
u_{L-1}(x)\\
u_{L+1}(x)
\end{array}\right)
\end{eqnarray}

\noindent exhibit a sinusoidal behavior in the asymptotic region, i.e. $x\rightarrow\infty$:

\begin{subequations}
\begin{eqnarray}\label{eq42a}
u_\ell(x) & \approx &  r_{\rm p}a^+\cos\left(r_{\rm p}x-L\pi/2 + \delta^+\right)
\\
\label{eq42b}
u_h(x) & \approx & -r_{\rm m}a^-\cos\left(r_{\rm m}x-L\pi/2 + \delta^-\right)
\end{eqnarray}
\end{subequations}

Given the definitions of $r_{\rm m}$ and $r_{\rm p}$ in Sec. II, we see that $u_\ell$ and $u_h$ are the light and heavy hole radial wavefunctions respectively.

\section{Born approximation}

\subsection{Green's matrix of the Schr\"odinger equation}

Inserting the expressions of $c^{\pm}(x)$ and $s^{\pm}(x)$ obtained from the integration of Eqs.~(\ref{eq35a}) and (\ref{eq35b}), in Eq.~(\ref{eq26}), yields the Lippmann-Schwinger integral equation in terms of partial waves \cite{Joachain}, satisfied by the two components $u_{L \pm 1}$ of the regular solution:

\begin{equation}\label{eq45}
\left( \begin{array}{c}
u_{L-1}(x)\\
u_{L+1}(x)\\
\end{array}\right)
= \left( \begin{array}{c}
u^0_{L-1}(x)\\
u^0_{L+1}(x)\\
\end{array}\right)
+ \int_0^{\infty} v(x')
\left( \begin{array}{cc}
{\mathcal G}_{L-1,L-1}^0(x,x') & {\mathcal G}_{L-1,L+1}^0(x,x')\\
{\mathcal G}_{L+1,L-1}^0(x,x') & {\mathcal G}_{L+1,L+1}^0(x,x')\\
\end{array}\right)
\left( \begin{array}{c}
u_{L-1}(x')\\
u_{L+1}(x')\\
\end{array}\right) {\rm d}x',
\end{equation}

\noindent where the four matrix elements ${\mathcal G}^0_{L \pm 1,L \pm 1}(x,x')$ are given in Appendix C. The functions $u^0_{L\pm 1}$ are the components of a regular solution of the Schr\"odinger equation of a free particle. They are obtained by taking $s^\pm = 0$ and $c^\pm = a_0^\pm$ in Eq.~(\ref{eq26}), where $a_0^+$ and $a_0^-$ are two constants that appear in the integration of the differential system, Eqs.~(\ref{eq35a}) and (\ref{eq35b}) (see Appendix A).

It is also possible to show that  the matrix $\left[{\mathcal G}^0(x,x')\right]$ whose elements are ${\mathcal G}_{L \pm 1,L \pm 1}^0(x,x')$ is a Green's matrix of the hole unperturbed Hamiltonian, $\left[H^0\right]$. Indeed, if one writes the Schr\"odinger equation, Eq.~(\ref{eq9}), as: $\left[H^0\right]\vec{u} = r_{\rm m}^2r_{\rm p}^2 v(x) \vec{u}$, one checks that the matrix $\left[{\mathcal G}^0\right]$ satisfies the following equation:

\begin{equation}\label{eq52}
\left[H^0\right]
\left[{\mathcal G}^0\right]
= r_{\rm m}^2r_{\rm p}^2~ \delta(x-x')
\left( \begin{array}{cc}
1 & 0\\
0 & 1\\
\end{array}\right),
\end{equation}

\noindent where the Dirac $\delta$-function comes from the differentiation of the step function $\Theta(x-x')$ that appears in the expression of the four matrix elements of $\left[{\mathcal G}^0\right]$ (see Appendix B).

\subsection{Born approximation}

The partial waves may be obtained in the Born approximation if the functions $u^0_{L\pm 1}(x')$ are substituted to the functions $u_{L\pm 1}(x')$ in the right hand side of the Lippmann-Schwinger equation, Eq.~(\ref{eq45}).

It is possible to find approximate expressions for the functions $\delta^{\pm}(x)$ and $a^{\pm}(x)$, by comparing $u^{\rm Born}_{L\pm 1}(x)$ to the general expression of $u_{L\pm 1}(x)$ given by Eq.~(\ref{eq26}). More precisely the identification of the terms in $\hat{n}_{L\pm 1}(r_{\rm p}x)$, $\hat{n}_{L\pm 1}(r_{\rm m}x)$, $\hat{\jmath}_{L\pm 1}(r_{\rm p}x)$ and $\hat{\jmath}_{L\pm 1}(r_{\rm m}x)$, gives the functions $\delta^+(x)$, $\delta^-(x)$, $a^+(x)$ and $a^-(x)$ respectively. Noting that $\delta^{\pm}\ll 1$ and keeping only terms of the first order in the potential strength yield:

\begin{subequations}
\begin{equation}\label{eq54a}
 \delta^+(x) \approx -r_{\rm p} \int_0^x v(x') \Big(\cos^2\frac{\displaystyle \alpha}{\displaystyle 2}~ \hat{\jmath}^2_{L-1}(r_{\rm p}x') + \sin^2\frac{\displaystyle \alpha}{\displaystyle 2}~ \hat{\jmath}^2_{L+1}(r_{\rm p}x')\Big) {\rm d}x' - r_{\rm m}~\frac{\displaystyle a_0^-}{\displaystyle a_0^+}~\int_0^x v(x') F(x') {\rm d}x'
\end{equation}
\begin{equation}\label{eq54b}
\delta^-(x) \approx -r_{\rm m} \int_0^x v(x') \Big(\sin^2\frac{\displaystyle \alpha}{\displaystyle 2}~ \hat{\jmath}^2_{L-1}(r_{\rm m}x') + \cos^2\frac{\displaystyle \alpha}{\displaystyle 2}~ \hat{\jmath}^2_{L+1}(r_{\rm m}x')\Big) {\rm d}x' - r_{\rm p}~\frac{\displaystyle a_0^+}{\displaystyle a_0^-}~\int_0^x v(x') F(x') {\rm d}x'
\end{equation}
\end{subequations}

\noindent where

\begin{eqnarray}\label{eq55}
F(x) = \sin\frac{\displaystyle \alpha}{\displaystyle 2}~\cos\frac{\displaystyle \alpha}{\displaystyle 2} \times \left[\hat{\jmath}_{L+1}(r_{\rm p}x)\hat{\jmath}_{L+1}(r_{\rm m}x) -\hat{\jmath}_{L-1}(r_{\rm p}x) \hat{\jmath}_{L-1}(r_{\rm m}x) \right] 
\end{eqnarray}

\noindent and

\begin{subequations}
\begin{eqnarray}\label{eq56a}
 a^+(x) &\approx& a_0^+\left[1-r_{\rm p} \int_0^x v(x') \Big(\cos^2\frac{\displaystyle \alpha}{\displaystyle 2}~ \hat{n}_{L-1}(r_{\rm m}x')\hat{\jmath}_{L-1}(r_{\rm p}x') + \sin^2\frac{\displaystyle \alpha}{\displaystyle 2}~ \hat{n}_{L+1}(r_{\rm p}x')\hat{\jmath}_{L+1}(r_{\rm p}x')\Big) {\rm d}x'\right]
 \nonumber\\
 &&{} - a_0^-r_{\rm m}~\sin\frac{\displaystyle \alpha}{\displaystyle 2}~\cos\frac{\displaystyle \alpha}{\displaystyle 2}\int_0^x v(x') \Big(\hat{n}_{L+1}(r_{\rm p}x')\hat{\jmath}_{L+1}(r_{\rm m}x') - \hat{n}_{L-1}(r_{\rm p}x')\hat{\jmath}_{L-1}(r_{\rm m}x')\Big) {\rm d}x'
\end{eqnarray}
\begin{eqnarray}\label{eq56b}
a^-(x) &\approx& a_0^-\left[1-r_{\rm m} \int_0^x v(x') \Big(\sin^2\frac{\displaystyle \alpha}{\displaystyle 2}~ \hat{n}_{L-1}(r_{\rm m}x')\hat{\jmath}_{L-1}(r_{\rm m}x') + \cos^2\frac{\displaystyle \alpha}{\displaystyle 2}~ \hat{n}_{L+1}(r_{\rm m}x')\hat{\jmath}_{L+1}(r_{\rm m}x')\Big) {\rm d}x'\right]
\nonumber\\
&&{} - a_0^+r_{\rm p}~\sin\frac{\displaystyle \alpha}{\displaystyle 2}~\cos\frac{\displaystyle \alpha}{\displaystyle 2}\int_0^x v(x') \Big(\hat{n}_{L+1}(r_{\rm m}x')\hat{\jmath}_{L+1}(r_{\rm p}x') - \hat{n}_{L-1}(r_{\rm m}x')\hat{\jmath}_{L-1}(r_{\rm p}x')\Big) {\rm d}x',
\end{eqnarray}
\end{subequations}

Since $\hat{n}_l(x) \approx -(2l-1)!!~x^{-l}$ and $\hat{\jmath}_l(x) \approx x^{l+1}/(2l+1)!!$ near the origin \cite{Calogero}, the integrals in Eqs.~(\ref{eq54a}), (\ref{eq54b}), (\ref{eq56a}) and (\ref{eq56b}), converge for potentials that vary like $x^{-\alpha}$ with $\alpha < 2$ near $x=0$. The presence of terms in $(a_0^+)^{-1}$ and $(a_0^-)^{-1}$ in $\delta^+$ and $\delta^-$ is not a problem since it is always possible to generate two independent solutions taking $a_0^+ \neq 0$ and $a_0^-\neq 0$. We note that since the phase shifts depend on the ratio $a_0^-/a_0^+$, which is arbitrary, there is no correlation between their sign and the attractive or repulsive nature of the potential. Therefore the meaning of the phase shifts $\delta^{\pm}$ is not as clear as it is for the phase shift $\delta_l$ of the partial waves when there is no coupling.
 
\section{Behavior of the wavefunctions near the origin}

The right hand side of Eq.~(\ref{eq37}) satisfied by the scattering amplitudes $a^{\pm}(x)$ and phase shifts $\delta^{\pm}(x)$, contains Neumann's functions $\hat{n}_l(x)$ that diverge when $x \rightarrow 0$. In some cases the reduced potential $v(x)$ may also diverge. Therefore, the computation of the solutions of Eq.~(\ref{eq37}) requires the knowledge of the behavior of either the functions $a^{\pm}(x)$ and $\delta^{\pm}(x)$ or the functions $c^{\pm}(x)$ and $s^{\pm}(x)$. Equations (\ref{eq35a}) and (\ref{eq35b}) show that this problem is equivalent to finding the approximate expressions of the wavefunctions $u_{L\pm 1}(x)$ valid near the origin.

One method is to solve the integral Lippmann-Schwinger equation, Eq.~(\ref{eq45}), by successive iterations. The most difficult case that may be treated in the present work is that of a potential $v(x)$ diverging in $x=0$ as a Coulomb potential, i.e. $v(x) \approx V_0x^{-1}$ when $x \rightarrow 0$. Singular potentials behaving like $x^{-m}$ with $m > $ 2, near the origin are not considered in the present work. After the first iteration, we see that all the integrals appearing in $u_{L\pm 1}^{(1)}(x)$ converge in $x=0$. However, in the general case, it is impossible to numerically solve the differential system Eq.~(\ref{eq37}) by substituting $u_{L\pm 1}^{(1)}(x)$ to $u_{L\pm 1}(x)$ near the origin. It is necessary to go to order 2 in the potential strength. After the second iteration, we note that one integral of $u_{L-1}^{(2)}(x)$ diverges logarithmically like $V_0^2 \int_0^x x'^{-1} {\rm d}x'$ unless $a_0^+$ and $a_0^-$ satisfy:

\begin{equation}\label{eq59}
a_0^+ \cos\frac{\displaystyle \alpha}{\displaystyle 2}~ r_{\rm p}^{L+1} = a_0^- \sin\frac{\displaystyle \alpha}{\displaystyle 2}~ r_{\rm m}^{L+1}
\end{equation}

\noindent If the above equality is not satisfied, it is not possible to solve the differential system Eq.~(\ref{eq37}) by imposing the values of the amplitudes at the origin, $a^{\pm}(0)$.

The general behavior of the regular solutions of the Schr\"odinger equation near the origin was studied by Newton in the case of a tensorial scattering potential, represented by a 2$\times$2 matrix \cite{Newton60}. For this kind of coupling the Green's matrix is diagonal. Using the Lippmann-Schwinger equation, it was found that one solution can be expanded as power series whereas the other exhibits an amplitude function that diverges logarithmically from the first iteration. Newton showed that one could circumvent this difficulty by modifying the inhomogeneity of the Lippmann-Schwinger equation, in our case $\left(\begin{array}{c}u_{L-1}^0\\u_{L+1}^0\end{array}\right)$, in a convenient way so that the problematic terms cancel each other out \cite{Newton60,Newton66}. According to Newton this method may always be applied in a general case, but the way to modify the integral equation may be more complicated either if the difference between the values of the two coupled angular momenta is greater than 2 or if the coupling is of a different nature.

The case of the scattering of holes is indeed very different: the potential is scalar, the Green's matrix is not diagonal and diverging terms appear only at the second iteration. This renders any use of Newton's method very difficult. For this reason we have adopted another method, which consists of taking the first order differential system, Eq.~(\ref{eq16}), satisfied by the vector $\left(\begin{array}{c} \vec{u}\\ \vec{G} \end{array}\right)$ defined in section III.A, as a starting point. The problem of logarithmic divergence that we found in the present problem, has actually its origin in the presence of a $x^{-1}$ term, called singularity of the first kind, in the expansion of the matrix $\left[A(x)\right] - \left[V(x)\right]$ near the origin. One result of the theory of differential systems is that the general regular solution generated by two arbitrary constants $(a,b)$ may be expanded as follows \cite{John65}:

\begin{equation}\label{eq66}
\left(\begin{array}{c} \vec{u}\\ \vec{G} \end{array}\right) = x^{\eta} \sum_i \sum_j \vec{C}_{i,j}(a,b) x^j \left[\log(x)\right]^i,
\end{equation}

\noindent where  $i$ and $j$ are integers that a priori satisfy: $0 \leq i \leq 3$ and $j \geq 0$; $\eta$ is an eigenvalue of the constant matrix $\left[A_0\right]$ that is the first term of the series expansion of the matrix $\left[A(x)\right]$ defined in Eq.~(\ref{eq17}): $\left[A(x)\right] = \sum_{n\geq 0} \left[A_n\right] x^{n-1}$. The  four-component constant vectors $\vec{C}_{i,j}(a,b)$ are determined recursively from the series expansion of $\left[A(x)\right]$. In the present work, the problem is reduced to the diagonalization and inversion of 4$\times$ 4 matrices.

The relationships between the constants $(a,b)$ and $(a_0^+,a_0^-)$ can be easily obtained from the comparison of the lowest order terms of the series expansions of $\vec{u}_{({\rm a,b})}$ and $u_{L\pm 1}^0(x)$. We found that that there exists one solution that may be expanded as a power series, generated by the couple $(a=0,b)$ \cite{BOG03}. At the lowest order the wavefunctions behave like: $u_{L\pm 1}(x) \approx x^{L+2}$. One can check that for this solution the equality in Eq.~(\ref{eq59}) is satisfied. If $a \neq 0$, the expansion contains a power series plus an additional term originating from the product of $V_0^2 \log(x)$ and the previous solution obtained with $(a=0,b=1)$. The $\log(x)$ term, known in the theory of Fuschian differential equations \cite{Ince26, CoddingtonLevinson55} appears only at the second order for potential behaving like a Coulomb potential near the origin, i.e. $v(x)\approx V_0 x^{-1}$ when $x\rightarrow 0$. The lowest term of the power series varies like $x^L$ for $u_{L-1}(x)$ and $V_0 x^{L+1}$ for $u_{L+1}(x)$. Note that the functions $u_{L\pm 1}$ are regular at $x=0$ for any couple $(a,b)$, despite the presence of a $\log(x)$ term in the expansion.

The functions $c^{\pm}(x)$ and $s^{\pm}(x)$, and hence $a^{\pm}(x)$ and $\delta^{\pm}(x)$, are deduced from $u_{L\pm 1}(x)$ and $G^{\pm}(x)$ solving Eq.~(\ref{eq26}) (we found that det$\left[W\right]=r_{\rm m}r_{\rm p}$). For any solution generated with $a \neq 0$, the constants $a_0^{\pm}$ do not correspond to the initial values of the amplitudes $a^{\pm}(x)$ since these functions diverge like $\log(x)$. They only coincide with the asymptotic values $a^{\pm}(\infty)$ if they can be calculated in the Born approximation for a very weak potential. For the solution that can be expanded as a power series, the constants $a_0^+$ and $a_0^-$ can be identified to the initial values of the amplitudes $a^+(0)$ and $a^-(0)$. In the Born approximation, the equalities $a^{\pm}(\infty) \simeq a_0^{\pm}$ are, of course, still valid.

The two independent solutions that we have just described exhibit the same behavior that we observed on the first iterations of the Lippmann-Schwinger equation at the beginning of this section. The power series appearing in the expansion of $\left(\begin{array}{c} \vec{u}\\ \vec{G} \end{array}\right)$ have a finite convergence radius $x_{\rm conv}$ that limits its use to a restricted interval; moreover $x_{\rm conv}$ is in general smaller than the point where the functions $a^{\pm}(x)$ and $\delta^{\pm}(x)$ cease to vary and which is directly linked to the range of the potential $v(x)$. To obtain $a^{\pm}(\infty)$ and $\delta^{\pm}(\infty)$, we thus solved the differential system Eq.~(\ref{eq37}) using the Runge-Kutta method from a value of $x$, $x = x_{\rm m} < x_{\rm conv}$.

\section{Numerical calculations}

Numerical calculations were performed to solve Eqs.~(\ref{eq37}) for the scattering phase shifts, amplitudes and radial wavefunctions. As examples we considered the scattering of the wave in the state $|F=3/2, L=1\rangle$, in Yukawa-type potentials frequently used to model charged point defects induced perturbation.

\subsection{The Yukawa potential}

The Yukawa potential we chose for the present example reads:

\begin{equation}\label{eq72}
v(x) = \frac{\displaystyle 2Z}{\displaystyle ka_{\rm B}}~\frac{\displaystyle e^{-x/k\lambda_{\rm s}}}{\displaystyle x},
\end{equation}

\noindent where $Z$ is the charge number, $a_{\rm B}$ the Bohr radius, and $\lambda_{\rm s}$ the screening length.

The potential above corresponds to the linearized Thomas-Fermi (LTF) description of the charge density devoted to the screening of a single ion. The LTF theory neglects the many-body effects contained in the Lindhard or Hubbard-Sham formalisms \cite{Greene}. The assumption of independent scattering by the various ions is valid only as long as the average defect separation is large compared to both the hole de Broglie wavelength and the screening length so that neighboring ion potentials do not overlap significantly \cite{Meyer2}.

In the Born approximation, the screening length $\lambda_{\rm s}$ is equal to the Dingle-Mansfield value: $
\lambda_{\rm s, Born}^{-2}= q^2/\epsilon_0 \epsilon_{\rm r} k_{\rm B}T~{\rm d}p/{\rm d}E_{\rm F}$ \cite{Dingle}, where $p$ is the hole density, $E_{\rm F}$ the Fermi energy and $\epsilon_{\rm r}$ the relative static dielectric constant. When the Born approximation is invalid, the energy dependence of the screening charge must be taken into account. In this case the screening length $\lambda_{\rm s}$ is evaluated numerically and self-consistently using the generalized Friedel sum rule \cite{Boardman} which insures that a given ion is fully screened at large distance.

Numerical calculations were performed taking the following values: $Z=-1$, $\lambda_{\rm s}=20a_{\rm B}$ with $a_{\rm B} = 4\pi\epsilon_0\epsilon_{\rm r}\hbar^2\gamma_1/m_0q^2$ and $\mu = 0.481$ for the effective mass parameter of Si \cite{Lipari}.

\subsubsection{Solutions without logarithmic divergence term}

\begin {figure}[!rh]
\centering
\scalebox{.33}{\rotatebox{0}{\includegraphics*{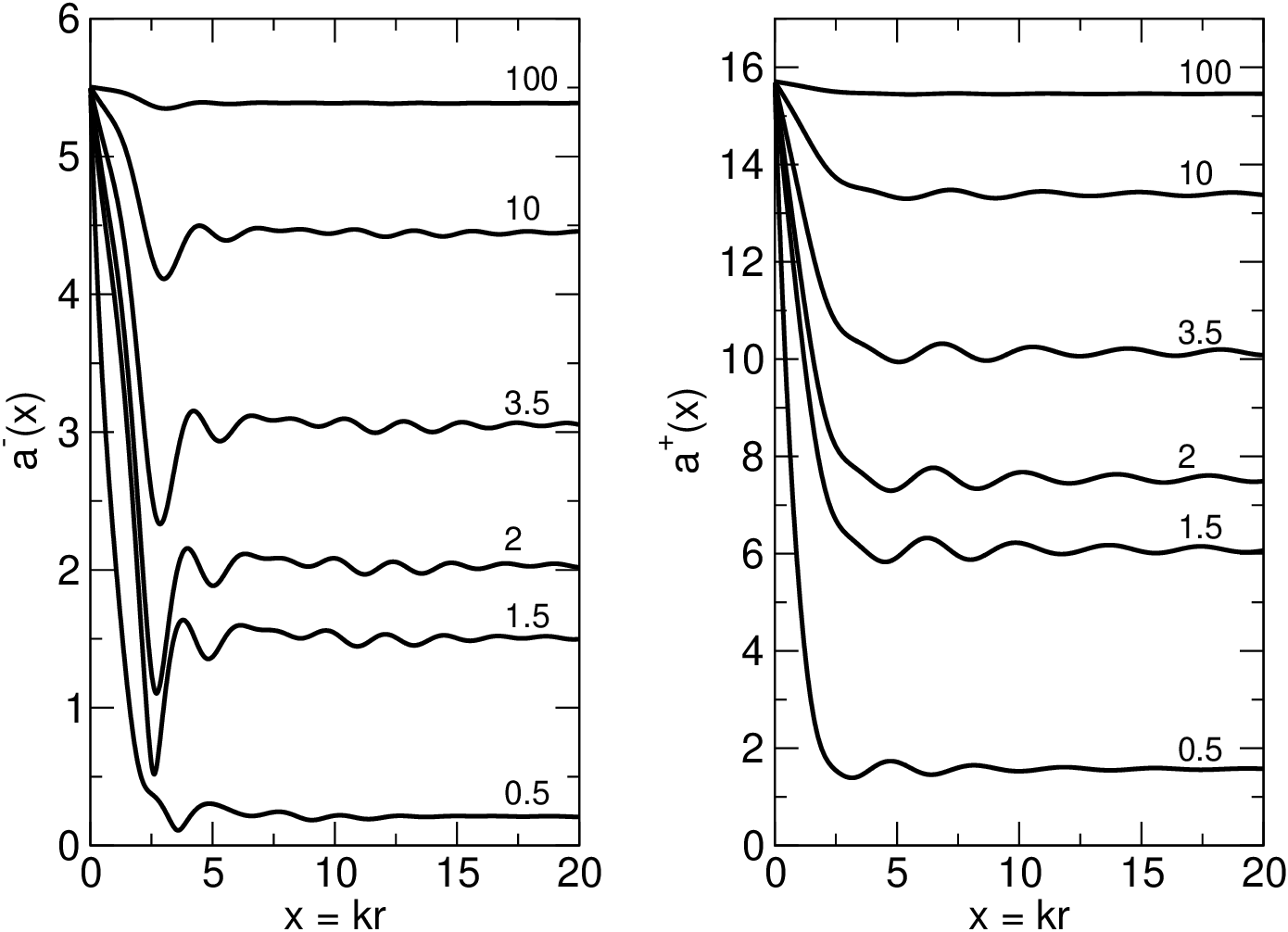}}}
\caption{Scattering amplitudes $a^-(x)$ and $a^+(x)$ as function of the scaled distance $x=kr$, for various values of $ka_{\rm B}$.}\label{fig1}
\end {figure}

\begin {figure}[!rh]
\centering
\scalebox{.33}{\rotatebox{0}{\includegraphics*{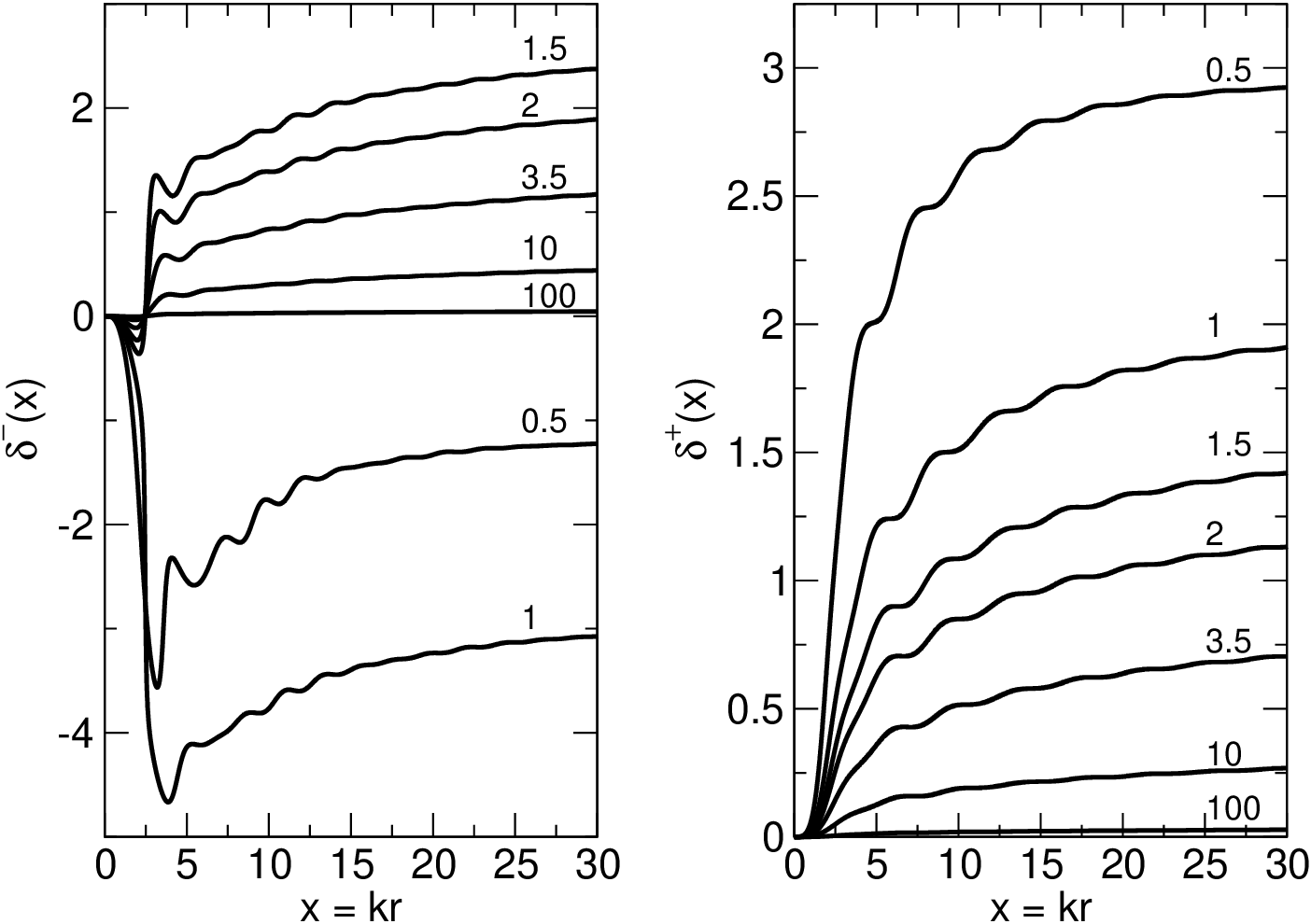}}}
\caption{Scattering phase shifts $\delta^-(x)$ and $\delta^+(x)$ as functions of the scaled distance $x=kr$, for various values of $ka_{\rm B}$.}\label{fig2}
\end {figure}

Scattering amplitudes and phase shifts of the solution without logarithmic divergence term are depicted in Figs.~\ref{fig1} and \ref{fig2}, for various values of the energy of the incoming hole, $E(k) = (ka_{\rm B})^2E_0$, where $E_0=\hbar^2\gamma_1/2m_0a_{\rm B}^2$ is the Rydberg energy. The curves were obtained by solving the differential system of Eq.~(\ref{eq37}) using the Runge-Kutta method, with $a^-(0) = a_0^- = (2L+3)!!/r_{\rm m}^{L+1}~ \cos (\alpha/2)$ and $a^+(0) = a_0^+ = (2L+3)!!/r_{\rm p}^{L+1}~ \sin(\alpha/2)$ taken as initial conditions (see Eq.~(\ref{eq59})). This choice corresponds to the following values: $a_0^-=5.505$ and $a_0^+=15.708$ for the $|F,L\rangle$ state that we consider.

Note that for $ka_{\rm B} \geq 100$ the phase shifts become very small and the amplitudes vary very little from their initial values. This is the regime where the Born approximation is valid and may be applied. Note that some of the phase shifts on Fig.~\ref{fig2} have not quite reached their asmptotic value for $x=30$. This is of no importance here since these quantities are not further exploited.

\begin {figure}[!rh]
\centering
\scalebox{.33}{\rotatebox{0}{\includegraphics*{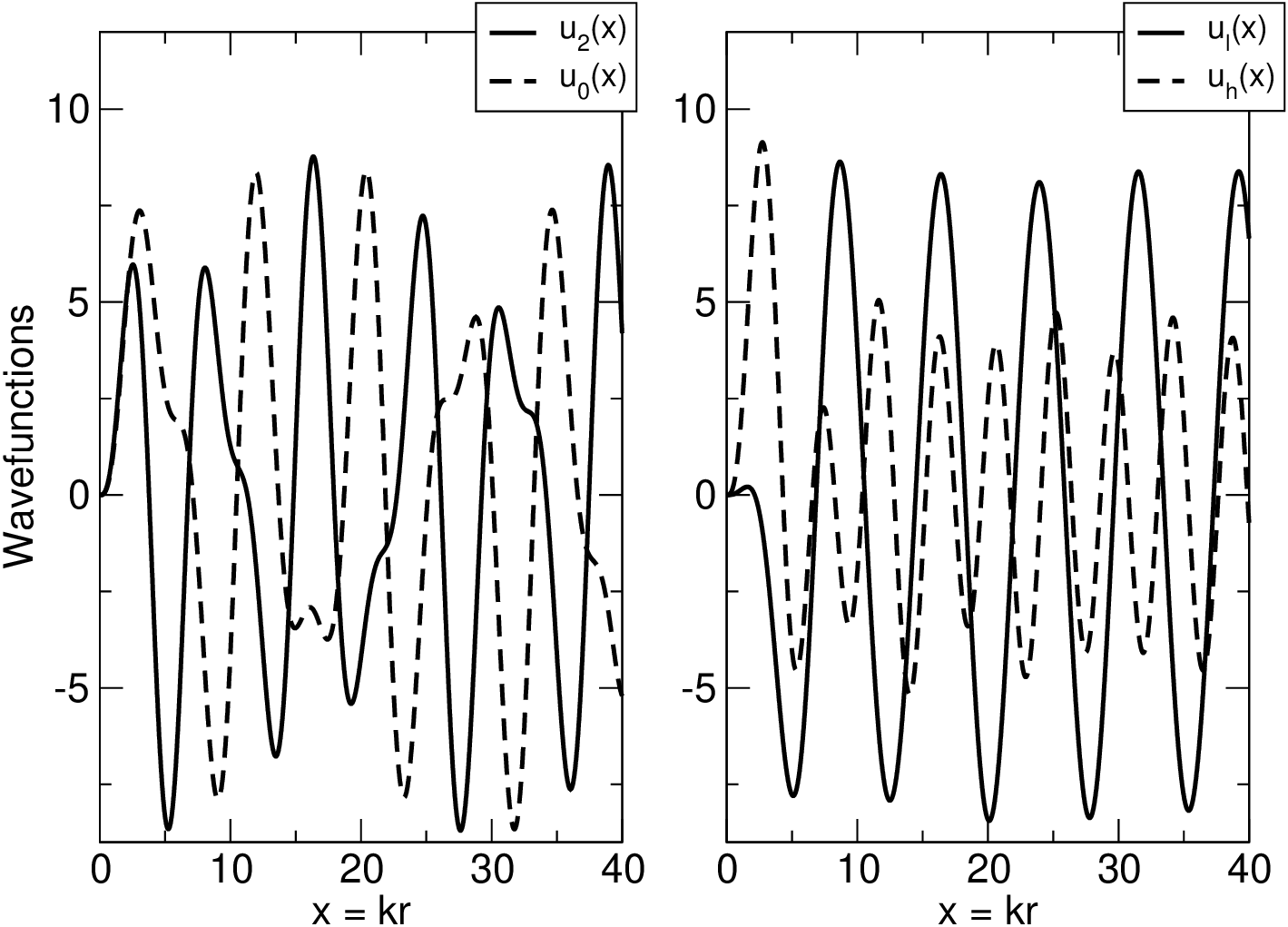}}}
\caption{Radial wavefunctions $u_0(x)$ and $u_2(x)$ (left) and light and heavy holes wavefunctions $u_\ell(x)$ and $u_{\rm h}$ (right) computed for $ka_{\rm B}=0.15$.}\label{fig3}
\end {figure}

The radial wavefunctions $u_0(x)$ and $u_2(x)$ as well as $u_\ell(x)$ and $u_{\rm h}(x)$ of the solution without the logarithmic divergence term in Fig.~\ref{fig3} were obtained from the scattering amplitudes and phase shifts discussed above.

One can check on Fig.~\ref{fig3} (right panel) that the light and heavy holes wavefunctions $u_\ell(x)$ and $u_{\rm h}(x)$ already start varying in a quasi-sinusoidal fashion when $x$ approaches $x=40$.

\begin {figure}[!rh]
\centering
\scalebox{.33}{\rotatebox{0}{\includegraphics*{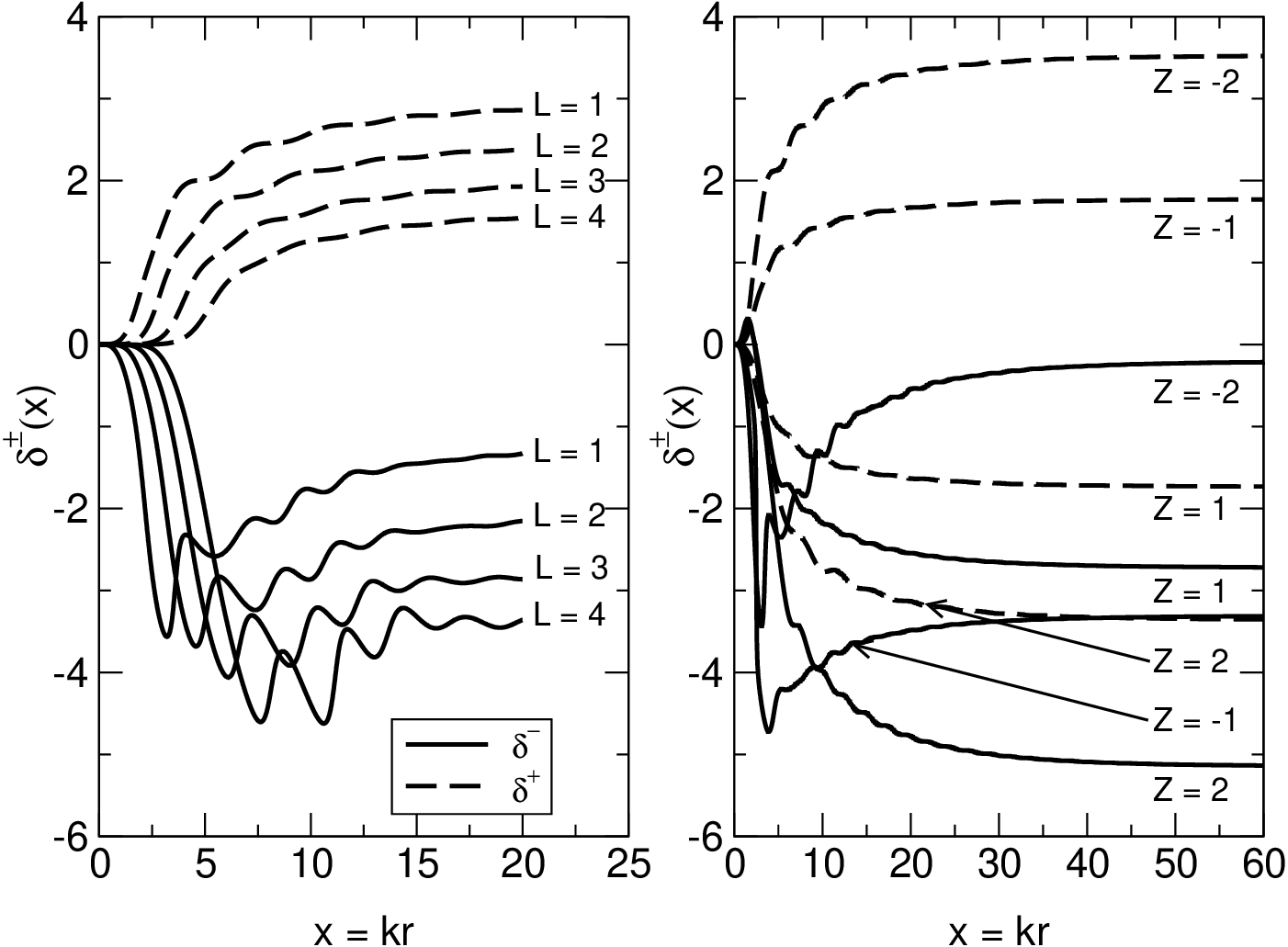}}}
\caption{Scattering phase shifts $\delta^-(x)$ (full line) and $\delta^+(x)$ (dashed line) as functions of $x$ for various values of orbital momentum $L$ (left panel) and charge number $Z$ (right panel).}\label{fig4}
\end {figure}

The influence of the orbital momentum $L$ and the charge number $Z$ of the defect on the phase shifts $\delta^{\pm}$ (solution without logarithmic diverging term) is depicted on Fig.~\ref{fig4}. Only the states $\xi=1$ for which $L=F-1/2$ were considered. We took the following values for the parameters: $Z=-1$, $\lambda_{\rm s}=20a_{\rm B}$ and $ka_{\rm B}=0.5$ for the study of the influence of $L$ on $\delta^{\pm}$ and $L=1$, $\lambda_{\rm s}=15a_{\rm B}$ and $ka_{\rm B}=1$ for the study of the influence of $Z$ on $\delta^{\pm}$.

One striking feature is that the phase shifts $\delta^+(x)$ and $\delta^-(x)$ behave very differently. The function $\delta^+(x)$ varies monotonically, is positive for attractive potentials and negative for repulsive potentials. Moreover its value decreases with increasing orbital momentum $L$. The function $\delta^+(x)$ has all the properties of the phase shifts of a scattering problem without coupling \cite{Calogero}. The phase shift $\delta^-(x)$ exhibits essentially two atypical behaviors: it is negative for attractive potentials ($Z=-1$ and $Z=-2$) and its behavior as a function of $L$ is ``normal'' only for small values of $x$. Indeed, outside the small $x$ region, $\delta^-(x)$ strongly oscillates before adopting a new trend: the absolute value of $\delta^-(x)$ increases with increasing orbital momentum $L$.

This unexpected behavior of $\delta^-(x)$ can be explained in the light of the study of the coupling terms $a^-D_{\rm c}/a^+$ appearing in the expression of the derivative of $\delta^+(x)$, ${\rm d}\delta^+(x)/{\rm d}x$, and $a^-D_{\rm c}/a^+$ appearing in the expression of ${\rm d}\delta^-(x)/{\rm d}x$ (see Eq.~(\ref{eq37})). The study of the ratio $a^+(x)/a^-(x)$ (not given here) shows clearly that, for a given value of $L$ that we take here to be $L=4$, it decreases for $x < 5$ before increasing suddenly, oscillating in the vicinity of $x \approx 10$ and finally converge towards 20. This coupling term has therefore a strong influence on the behavior of $\delta^-(x)$, while it does not affect $\delta^+(x)$ significantly. 

\subsubsection{Solutions including logarithmic divergence term}

\begin {figure}[!rh]
\centering
\scalebox{.33}{\rotatebox{0}{\includegraphics*{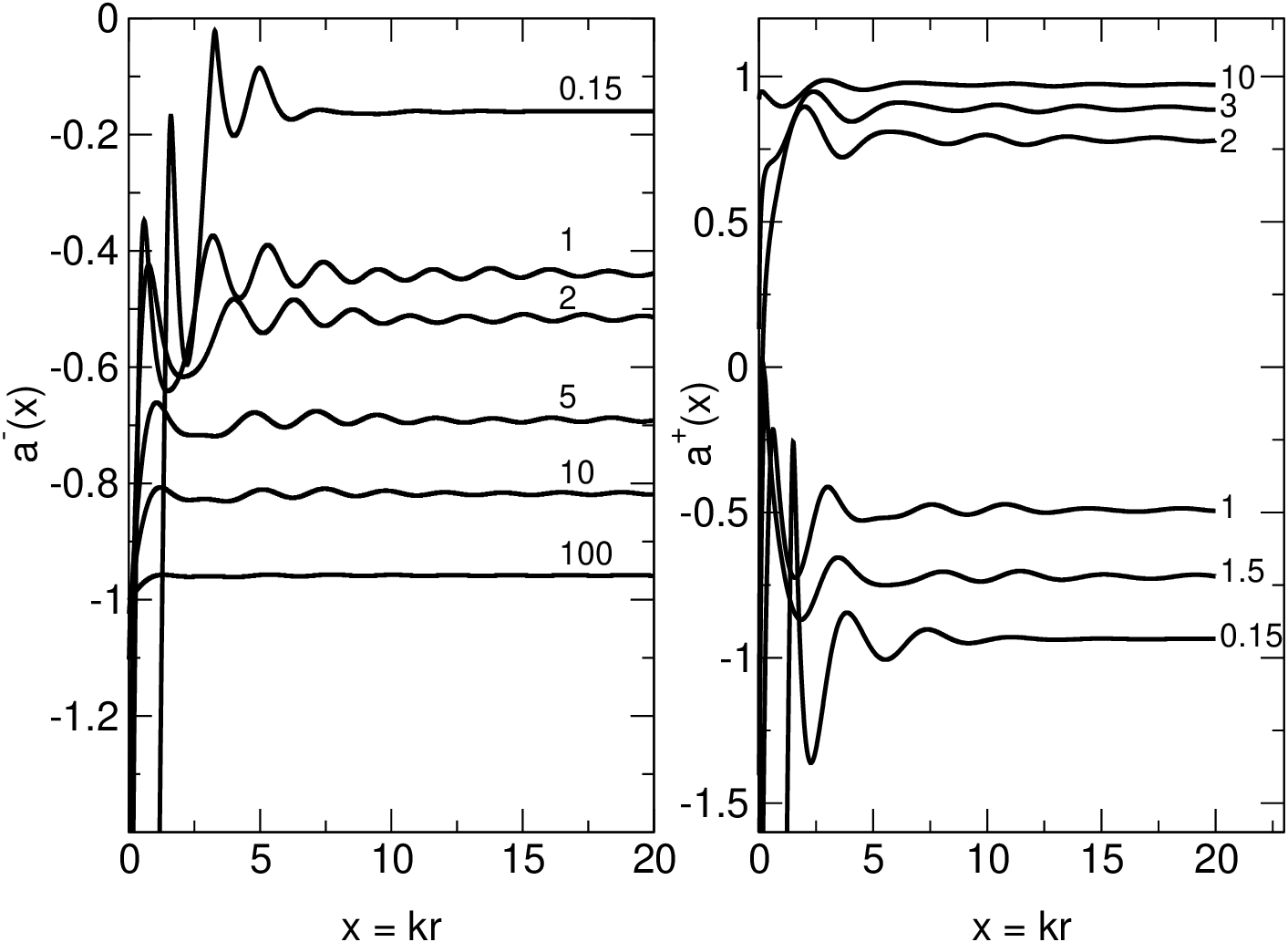}}}
\caption{Scattering amplitudes as functions of $x$ considering the logarithmic divergence term, for various values of $ka_{\rm B}$.}\label{fig5}
\end {figure}

Another solution whose amplitudes presents a logarithmic divergence, was generated. We chose the following values for the integration constants: $a_0^-=-1$ and $a_0^+=1$, and kept the previous parameters: $\lambda_{\rm s}=20~a_{\rm B}$, $Z=-1$ and $L=1$. The logarithmic divergence of the amplitudes near $x=0$ appears very well on Fig.~\ref{fig5}. For large values of the energy of the hole, $E(k)$ ($ka_{\rm B} > 100$ for $a^-$ and $ka_{\rm B}>10$ for $a^+$), the asymptotic values, $a^{\pm}(\infty)$ and the parameters $a_0^{\pm}$ are practically the same. 

\begin {figure}[!rh]
\centering
\scalebox{.33}{\rotatebox{0}{\includegraphics*{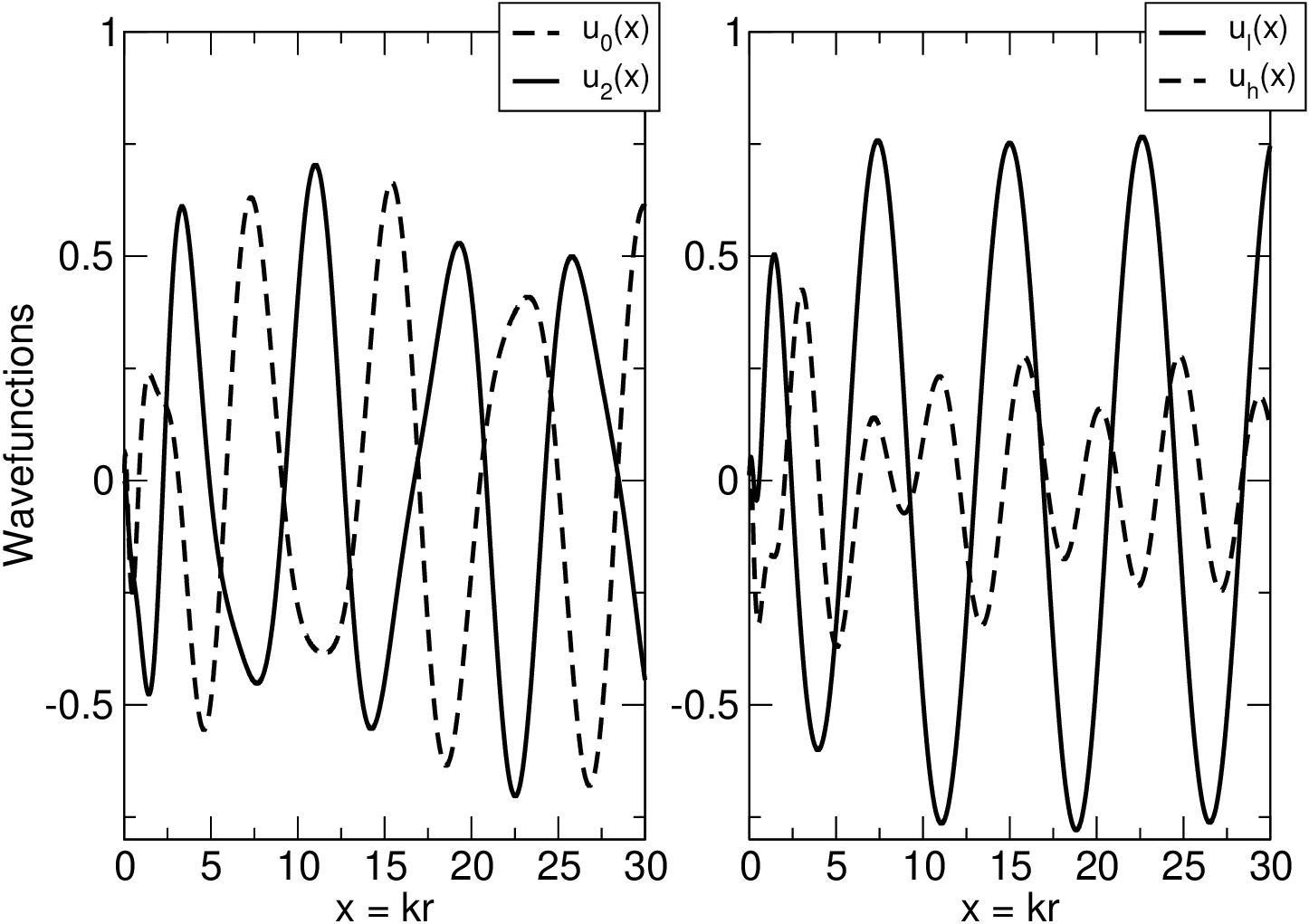}}}
\caption{Wavefunctions $u_0$, $u_2$, $u_\ell$ and $u_{\rm h}$ as functions of $x$.}\label{fig6}
\end {figure}

The radial wavefunctions $u_0(x)$, $u_2(x)$, $u_\ell(x)$ and $u_{\rm h}(x)$, obtained for $ka_{\rm B}=0.15$ are depicted in Fig.~\ref{fig6}.

One can check that the logarithmic divergence does not compromise the regularity of the radial wavefunctions near $x=0$ and that $u_\ell(x)$ and $u_{\rm h}(x)$ start exhibiting sinusoidal oscillations from $x\approx 30$.

\subsection{Modified Yukawa potential}

The last case treated here is that of the scattering of the wave $|F=3/2,L=1\rangle$ by a repulsive Yukawa potential ($Z = 1$, $\lambda_{\rm s} = 200~a_{\rm B}$) modified by an attractive square core ($V = -1.6 E_0$ for $r<r_1 = 10~a_{\rm B}$). This modified Yukawa potential has a positive maximum $V_{\rm max} \approx 0.19~E_0$.

Because of its rapid variations, this type of potential is not rigorously compatible with the use of the effective mass approximation. It is nonetheless frequently used to model the chemical shift that is observed on the bound states of doping impurities \cite{ELG80}. It is not in the scope of the present work to study the effect of the core potential on the phase shifts $\delta^{\pm}$, but it is of interest to put in evidence the influence of the energy $E(k)$ on the radial wavefunctions $u_{L\pm 1}(x)$ in the case of a non-monotonical potential. Since the perturbation potential is regular in $x=0$ ($V_0=0$), the amplitudes $a^{\pm}$ do not exhibit a logarithmic divergence and the two independent solutions, $({\rm a})$ and $({\rm b})$, can be expanded as a power series.

\begin {figure}[!rh]
\centering
\scalebox{.33}{\rotatebox{0}{\includegraphics*{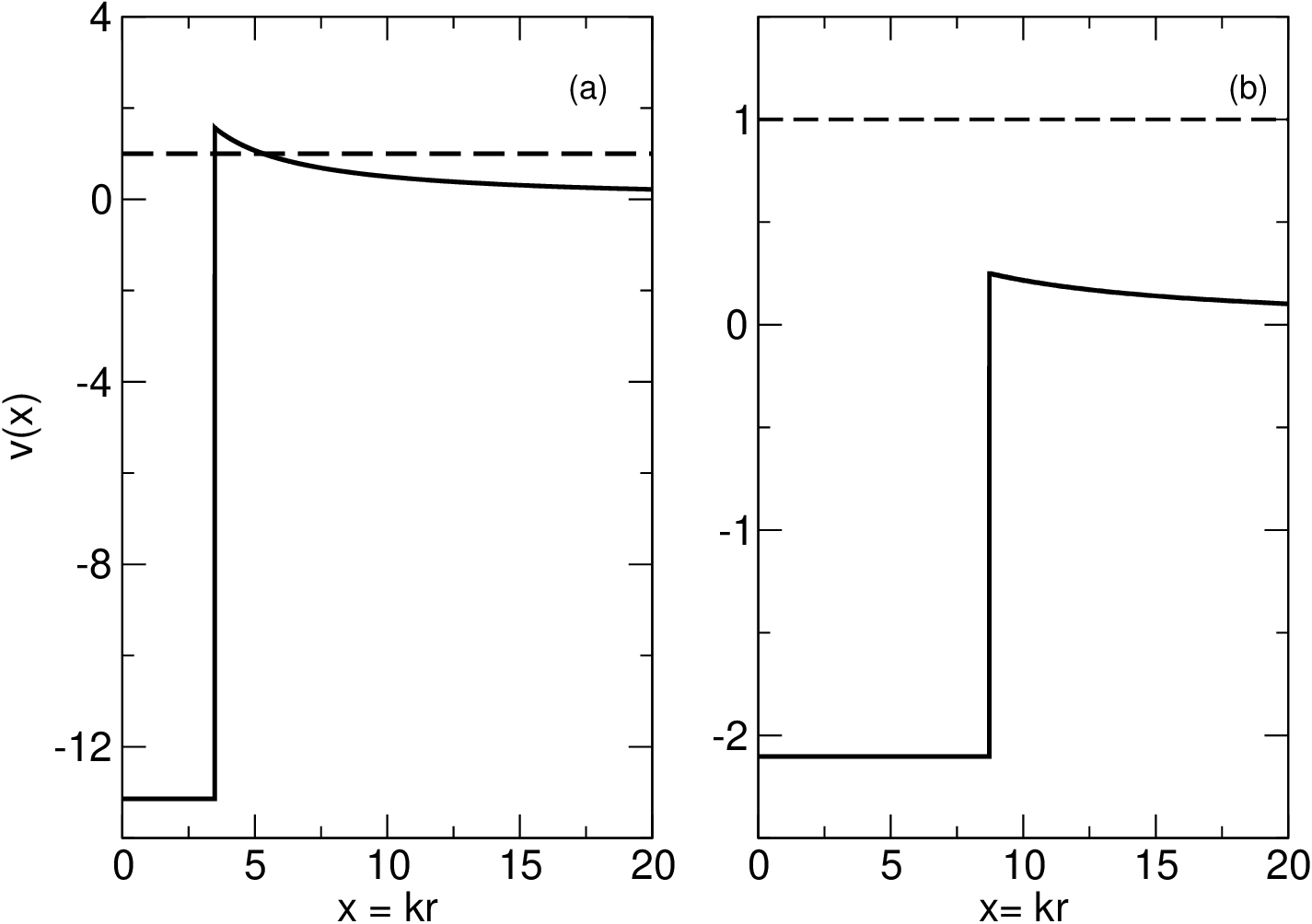}}}
\caption{Modified Yukawa potentials as functions of $x$. The dotted lines indicate the location of energy of the hole with respect to the potential barriers in full lines. Two cases are considered: $v(x) = V/E(k)$ for $E(k) = 0.64~V_{\rm max}$ on figure $({\rm a})$ and $E(k) = 4~V_{\rm max}$ on figure $({\rm b})$.}\label{fig7}
\end {figure}

The behavior of the scattered waves is different if the energy of the incoming hole, $E(k)$, is greater or smaller than $V_{\rm max}$. We chose two values for the hole energy: $E(k) = 0.64~V_{\rm max}$ and $E(k) = 4~V_{\rm max}$, that corresponds to $ka_{\rm B}=0.349$ and $ka_{\rm B}=0.872$ respectively. The reduced potentials and hole energies are depicted in Fig.~(\ref{fig7}).

The radial wavefunctions $u_0(x)$ and $u_2(x)$ are depicted on Figs.~8 and 9. They were obtained by taking the couple $(a^-(0)= a_0^- = \cos\frac{\displaystyle \alpha}{\displaystyle 2}~/r_{\rm m}^{L+1} = 0.367,~a^+(0) = a_0^+ = \sin\frac{\displaystyle \alpha}{\displaystyle 2}~/r_{\rm p}^{L+1}=1.047)$ for the solution $({\rm a})$ and $(a^-(0) = a_0^- =-1,~a^+(0) = a_0^+ = 1)$ for the solution $({\rm b})$ as initial conditions. 

\begin {figure}[!rh]
\centering
\scalebox{.33}{\rotatebox{0}{\includegraphics*{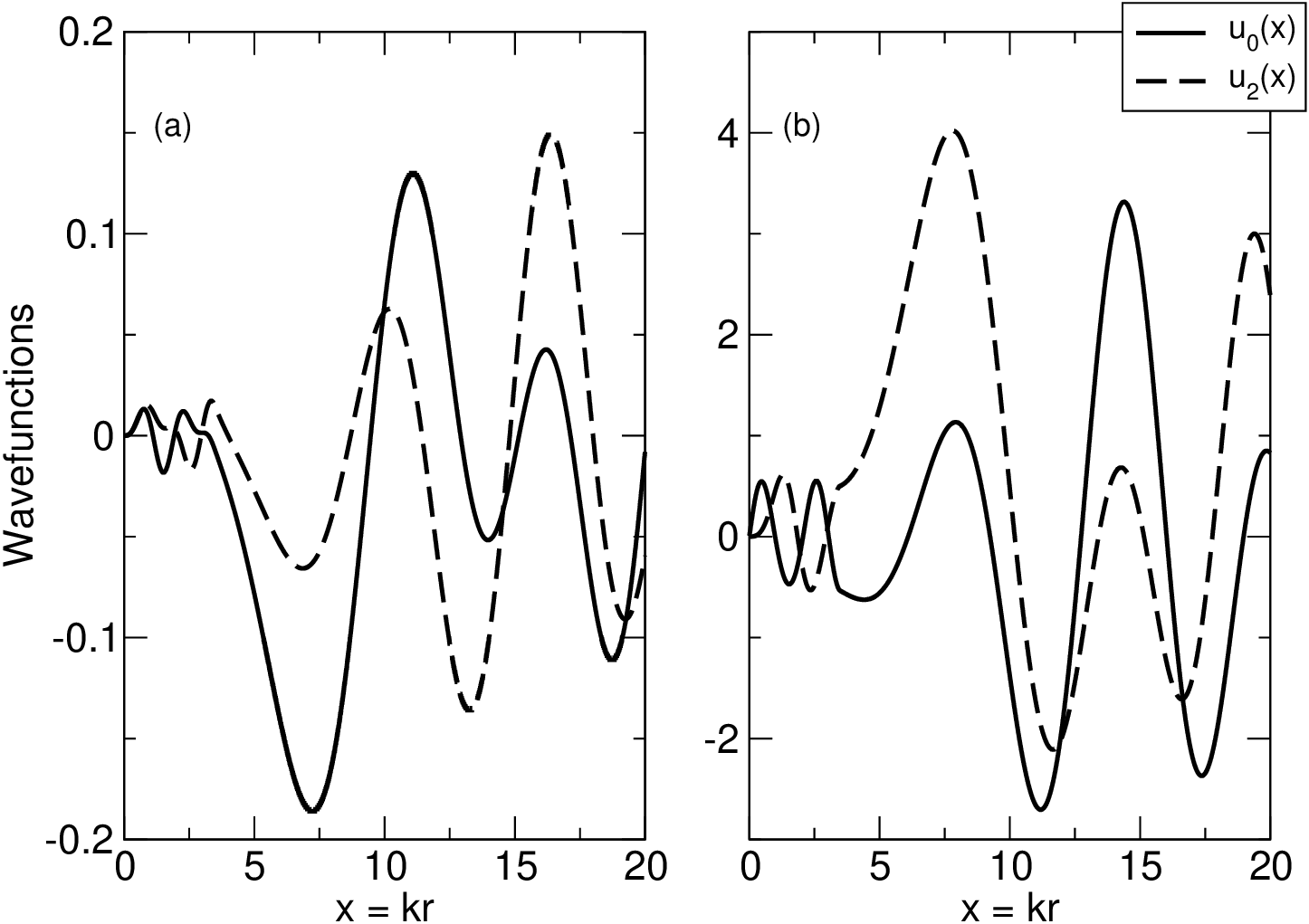}}}
\caption{Radial wavefunctions $u_0(x)$ and $u_2(x)$ of the solutions $({\rm a})$ and $({\rm b})$ for $E(k)=0.64~V_{\rm max}$.}\label{fig8}
\end {figure}

For $E(k) < V_{\rm max}$ the vanishing wave that goes through the classically forbidden region, $\left[3.5;5.5\right]$, by tunnel effect can be seen on the radial wavefunctions $u_0(x)$ and $u_2(x)$ of the solution $({\rm b})$, in Fig.~\ref{fig8}. The high value of $r_1$ allows one to clearly see the rapid oscillations in the region where the potential is attractive.

\begin{figure}[!rh]
\centering
\scalebox{.33}{\rotatebox{0}{\includegraphics*{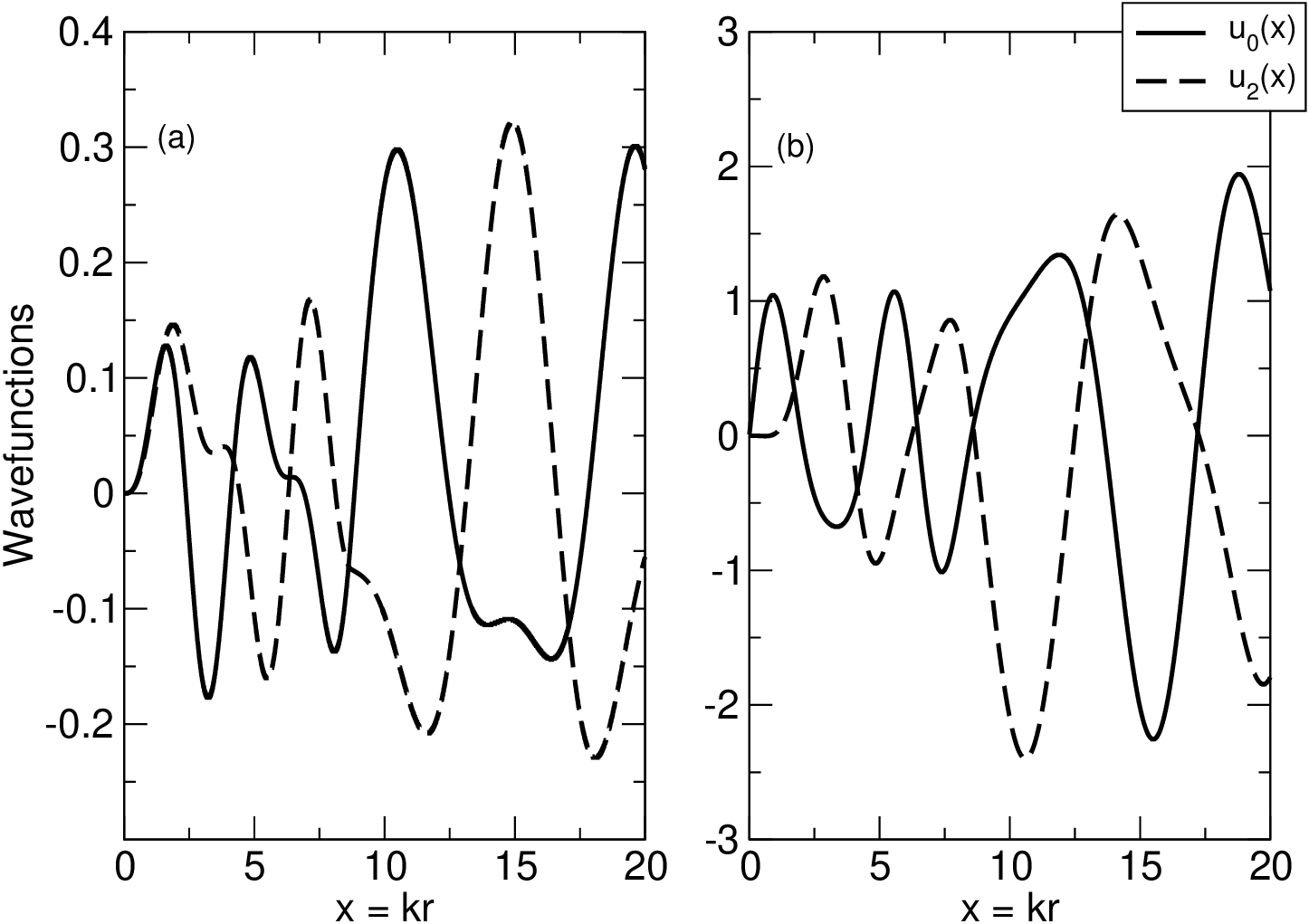}}}
\caption{Radial wavefunctions $u_0(x)$ and $u_2(x)$ of the solutions $({\rm a})$ and $({\rm b})$ for $E(k)=4~V_{\rm max}$.}\label{fig9}
\end{figure}

For $E(k) > V_{\rm max}$ (Fig.~{\ref{fig9}) the wavefunctions exhibit an oscillatory behavior for all the values of $x$ and their amplitudes are of the same order of magnitude both in the well region and the barrier.

\section{Summary, discussion and Conclusion}

In this article we addressed the problem of realistic calculations of scattering states of holes in mixed valence bands in the strong spin-orbit coupling regime. From the early stage of the present work it appeared that the generalization of Ralph's results to finite range potentials, such as those created by ionized defects, was indeed specially involved. In particular, the behavior of the two radial wavefunctions $u_{L\pm 1}(x)$ near the origin required special attention. We found that the variable phase method provided us with a very convenient framework that facilitated such calculations.

Starting from the two coupled Schr\"odinger equations satisfied by the radial wavefunctions, and introducing auxiliary functions, we first derived the system of four coupled non-linear differential equations whose solutions are the two phases and two amplitudes characterizing the scattering states with well defined total angular momentum and parity $|F,L\rangle$. This systems forms the generalized phase equations \cite{Calogero}. Ralph's parameters were obtained by integrating these equations over an interval larger than the range of the perturbation potential. Then we obtained an expression of the Green's matrix of the unperturbed Schr\"odinger equation as well as the Lippmann-Schwinger integral equation which yields an iterative scheme to evaluate the radial wavefunctions. Considering only the first iteration, we found the ``Born approximation'' for the amplitudes and phases of the scattering states.

Owing to the non-diagonal form of the Green's matrix, the expansions of the solutions near the origin ($x \approx 0$) that are necessary for successful numerical integration of the generalized phase equation, have been constructed by making use of a technique only applicable to first order differential systems \cite{John65}. We showed that for Coulomb-like potentials, there exists logarithmically diverging terms in the amplitudes near the origin, appearing only at the second order in the potential strength. This specific result stems from the coupling of the holes wavefunctions: for the tensorial couplings that we found in the litterature \cite{Newton60,Cox65,Newton66}, the $\log (x)$ terms are indeed generated from the first iteration of the Born series. This fact might explain why we could not find published results on the problem we treat in the present paper since Ralph's article \cite{Ralph}.

The generalized phase equations were solved for screened Coulomb potentials of the Yukawa type and various values of angular momentum, incoming hole energy and defect charge state. The oscillatory but non-sinusoidal behavior of the radial wavefunctions $u_{L\pm 1}(x)$ in the asymptotic region is coherent with the predictions of Ralph's theory \cite{Ralph}. By linear combination of those functions, we constructed the heavy and light hole wavefunctions which  exhibit a sinusoidal asymptotic behavior. This result shows that the phase shifts, solutions of the generalized phase equations, are those of the heavy and light holes. However their meaning is not as clear as for the phase shifts $\delta_l$ of the partial waves in the absence of coupling since their signs do not only depend on the charge of the ionized defect but also on the choice of two arbitrary integration constants.

Calculations of Hall and drift mobilities require proper computation of the energy average of the various relaxation time functions $\tau_{ij}$ \cite{Ralph,COS74,SZM86}. A study of the influence of the temperature and concentration of doping atoms on those physical quantities is practically achievable only on the condition that the variations of the relaxation times $\tau_{ij}$, over a range of incoming hole energy $E$ and screening length $\lambda_{\rm s}$ that is accessible to experiments, could be reproduced by approximate analytic forms analogous to those of Meyer and Bartoli \cite{Meyer}. A vast amount of purely numerical work remains to be done to construct those approximate analytic expressions because it may appear necessary to consider a large number of scattering states $|F,L\rangle$ for some values of $(E,\lambda_{\rm s})$. This work is beyond the scope of the present article.

The strong spin-orbit coupling regime is well adapted to semiconductors like Ge and GaAs, where the energy of spin-orbit interaction is about $\Delta_{\rm SO}\approx 300$ meV. The study of the scattering states in the weak coupling regime, i.e. $\Delta_{\rm SO}\rightarrow 0$, can be seen as the other limit case that can be tackled before addressing the more involved scattering problems in Si for which $\Delta_{\rm SO} = 40$ meV, i.e. well in the intermediate coupling regime. The variable phase method can be used again since the two radial wavefunctions $u_{L\pm 1}$ are solutions of a differential system of the second order analogous to the one in the strong coupling regime \cite{Baldereschi1}.

\appendix

\section{Free solution}

The 4$\times$4 matrix $\left[W(x)\right]$ whose columns are the four linearly independent free solutions ($v=0$) of the differential system Eq.~(\ref{eq16}), reads:

\begin{eqnarray}\label{eqa1}
\left[W\right]=
\left(\begin{array}{cccc}
\vec{u}^+_{0,\rm r} &\vec{u}^+_{0,\rm i} &\vec{u}^-_{0,\rm r} &\vec{u}^-_{0,\rm i} \\
\vec{G}^+_{0,\rm r} &\vec{G}^+_{0,\rm i} &\vec{G}^-_{0,\rm r} &\vec{G}^-_{0,\rm i} \\
\end{array}\right),
\end{eqnarray}

\noindent where

\begin{subequations}
\begin{eqnarray}\label{eqa2}
\vec{G}^+_{0,{\rm r}}=
\left(\begin{array}{c}
\hat{\jmath}_L(r_{\rm p}x)\\
0\\
\end{array}\right)~\mbox{and}~
\vec{G}^-_{0,{\rm r}}=
\left(\begin{array}{c}
0\\
\hat{\jmath}_L(r_{\rm m}x)\\
\end{array}\right),\\
\vec{G}^+_{0,{\rm i}}=
\left(\begin{array}{c}
\hat{n}_L(r_{\rm p}x)\\
0\\
\end{array}\right)~\mbox{and}~
\vec{G}^-_{0,{\rm i}}=
\left(\begin{array}{c}
0\\
\hat{n}_L(r_{\rm m}x)\\
\end{array}\right),
\end{eqnarray}
\end{subequations}

\noindent and

\begin{subequations}
\begin{eqnarray}\label{eqa3}
\vec{u}^+_{0}=
\left(\begin{array}{c}
r_{\rm p}\cos\frac{\displaystyle \alpha}{\displaystyle 2}~ \hat{z}_{L-1}(r_{\rm p}x)\\
\rule{0pt}{4ex}r_{\rm p}\sin\frac{\displaystyle \alpha}{\displaystyle 2}~ \hat{z}_{L+1}(r_{\rm p}x)\\
\end{array}\right)
\end{eqnarray}
\begin{eqnarray}
\vec{u}^-_{0}=
\left(\begin{array}{c}
-r_{\rm m}\sin\frac{\displaystyle \alpha}{\displaystyle 2}~ \hat{z}_{L-1}(r_{\rm m}x)\\
\rule{0pt}{4ex}r_{\rm m}\cos\frac{\displaystyle \alpha}{\displaystyle 2}~ \hat{z}_{L+1}(r_{\rm m}x)\\
\end{array}\right),
\end{eqnarray}
\end{subequations}

\noindent where $\hat{\jmath}_l$ and $\hat{n}_l$ are the modified spherical Bessel and Neumann functions respectively \cite{Calogero} and $\hat{z}_l =\hat{\jmath}_l$ for the regular solutions and $\hat{z}_l =\hat{n}_l$ for the irregular solutions.

Taking $\vec{C}_0 = \left(\begin{array}{c} a_0^+\\ 0\\ a_0^-\\ 0 \end{array}\right)$ in Eq.~(\ref{eq25}) yields the free regular radial wavefunctions that appear in the Lippmann-Schwinger equation, Eq.~(\ref{eq45}):

\begin{subequations}
\begin{equation}\label{eqa4}
u^0_{L-1}(x) = r_{\rm p}\cos \frac{\displaystyle \alpha}{\displaystyle 2}~ a_0^+\hat{\jmath}_{L-1}(+) - r_{\rm m} \sin \frac{\displaystyle \alpha}{\displaystyle 2}~ a_0^-\hat{\jmath}_{L-1}(-)
\end{equation}
\begin{equation}\label{eqa5}
u^0_{L+1}(x) = r_{\rm p}\sin \frac{\displaystyle \alpha}{\displaystyle 2}~ a_0^+\hat{\jmath}_{L+1}(+) + r_{\rm m} \cos \frac{\displaystyle \alpha}{\displaystyle 2}~ a_0^-\hat{\jmath}_{L+1}(-)
\end{equation}
\end{subequations}

\section{Functions $D^{\pm}$, $SD^{\pm}$, $D_{\rm c}$ and $SD^{\varepsilon,-\varepsilon}_{\rm c}$}

Below are given the various functions that appear in the generalized equations for the phases and amplitudes, Eq.~(\ref{eq37}):

\begin{equation}\label{eq1b}
D^{\pm} = \cos^2\frac{\displaystyle \alpha}{\displaystyle 2}~\left[D^{\pm}_{L\mp 1}\right]^2 + \sin^2\frac{\displaystyle \alpha}{\displaystyle 2}~\left[D^{\pm}_{L\pm 1}\right]^2,
\end{equation}
\begin{equation}\label{eq2b}
SD^{\pm} = \cos^2\frac{\displaystyle \alpha}{\displaystyle 2}~S^{\pm}_{L\mp 1}D^{\pm}_{L\mp 1} + \sin^2\frac{\displaystyle \alpha}{\displaystyle 2}~S^{\pm}_{L\pm 1}D^{\pm}_{L\pm 1},
\end{equation}
\begin{equation}\label{eq3b}
D_{\rm c} = D^+_{L-1}D^-_{L-1} - D^+_{L+1}D^-_{L+1},
\end{equation}
\begin{equation}\label{eq4b}
SD^{\varepsilon,-\varepsilon}_{\rm c} = \sin\frac{\displaystyle \alpha}{\displaystyle 2}~\cos\frac{\displaystyle \alpha}{\displaystyle 2}~\left(S^{\varepsilon}_{L-1}D^{-\varepsilon}_{L-1} - S^{\varepsilon}_{L+1}D^{-\varepsilon}_{L+1}\right),
\end{equation}
\noindent with $\varepsilon=\pm$,
\begin{equation}\label{eq5b}
D^{\pm}_l = \cos(\delta^{\pm}) \hat{\jmath}_l(\pm) - \sin(\delta^{\pm})\hat{n}_l(\pm),
\end{equation}
\noindent and
\begin{equation}\label{eq6b}
S^{\pm}_l = \cos(\delta^{\pm}) \hat{n}_l(\pm) + \sin(\delta^{\pm})\hat{\jmath}_l(\pm).
\end{equation}

For ease of notation in Eqs.~(\ref{eq5b}) and (\ref{eq6b}), the arguments $r_{\rm m}x$ and $r_{\rm p}x$ of the modified spherical Bessel and Neumann functions are denoted $-$ and $+$ respectively.

\section{Green's matrix elements}

The expressions of the four matrix elements of the Green's matrix $\left[{\mathcal G}^0\right]$ defined in Eq.~(\ref{eq45}):

\begin{equation}\label{eq1c}
{\mathcal G}_{L-1,L-1}^0(x,x') = r_{\rm p}\cos^2 \frac{\displaystyle \alpha}{\displaystyle 2}~ g_{L-1}(r_{\rm p}x,r_{\rm p}x') + r_{\rm m}\sin^2 \frac{\displaystyle \alpha}{\displaystyle 2}~g_{L-1}(r_{\rm m}x,r_{\rm m}x')
\end{equation}
\begin{equation}\label{eq2c}
{\mathcal G}_{L+1,L+1}^0(x,x') = r_{\rm p}\sin^2 \frac{\displaystyle \alpha}{\displaystyle 2}~ g_{L+1}(r_{\rm p}x,r_{\rm p}x') + r_{\rm m}\cos^2 \frac{\displaystyle \alpha}{\displaystyle 2}~g_{L+1}(r_{\rm m}x,r_{\rm m}x')
\end{equation}
\begin{eqnarray}\label{eq3c}\nonumber
{\mathcal G}_{L-1,L+1}^0(x,x') & = &\sin \frac{\displaystyle \alpha}{\displaystyle 2}~\cos \frac{\displaystyle \alpha}{\displaystyle 2}~ \left[r_{\rm p} \Big(\hat{\jmath}_{L+1}(r_{\rm p}x') \hat{n}_{L-1}(r_{\rm p}x) - \hat{\jmath}_{L-1}(r_{\rm p}x) \hat{n}_{L+1}(r_{\rm p}x')\Big) \right. \\
& & -\left. r_{\rm m} \Big( \hat{\jmath}_{L+1}(r_{\rm m}x') \hat{n}_{L-1}(r_{\rm m}x) - \hat{\jmath}_{L-1}(r_{\rm m}x) \hat{n}_{L+1}(r_{\rm m}x') \Big) \right] \Theta(x-x')
\end{eqnarray}
\begin{eqnarray}\label{eq4c}\nonumber
{\mathcal G}_{L+1,L-1}^0(x,x') & = &\sin \frac{\displaystyle \alpha}{\displaystyle 2}~\cos \frac{\displaystyle \alpha}{\displaystyle 2}~ \left[r_{\rm p} \Big(\hat{\jmath}_{L-1}(r_{\rm p}x') \hat{n}_{L+1}(r_{\rm p}x) - \hat{\jmath}_{L+1}(r_{\rm p}x) \hat{n}_{L-1}(r_{\rm p}x')\Big) \right. \\
& & -\left. r_{\rm m} \Big( \hat{\jmath}_{L-1}(r_{\rm m}x') \hat{n}_{L+1}(r_{\rm m}x) - \hat{\jmath}_{L+1}(r_{\rm m}x) \hat{n}_{L-1}(r_{\rm m}x') \Big) \right] \Theta(x-x'),
\end{eqnarray}

\noindent where $\Theta(x)$ is the step function and 

\begin{equation}\label{eq5c}
g_l(x,x') = \Big[\hat{\jmath}_l(x') \hat{n}_l(x) - \hat{\jmath}_l(x) \hat{n}_l(x')\Big]\times \Theta(x-x').
\end{equation}

\bibliography{PRB_Ref}

\end{document}